\documentclass[a4paper,fleqn,usenatbib]{mnras}

\usepackage{newtxtext,newtxmath}

\usepackage[T1]{fontenc}
\usepackage{ae,aecompl}

\usepackage{graphicx}	
\usepackage{amsmath}	
\usepackage{xcolor}		

\usepackage{gensymb} 

\title[A pair of Jovian Trojans]{A pair of Jovian Trojans at the L4 Lagrange point}

\author[T. R. Holt et al.]{Timothy R. Holt,$^{1,2}$\thanks{E-mail: timothy.holt@usq.edu.au (TRH)}
 David Vokrouhlick{\'y},$^{3}$ David Nesvorn{\'y},$^{2}$ Miroslav Bro\v{z}$^{3}$
 and \newauthor Jonathan Horner$^{1}$
\\
$^{1}$Centre for Astrophysics, University of Southern Queensland, Toowoomba, Queensland 4350, Australia\\
$^{2}$Southwest Research Institute, Department of Space Studies, 1050 Walnut St., Ste 300, Boulder, CO-80302, USA\\
$^{3}$Institue of Astronomy, Charles University, V Hole\v{s}ovi\v{c}k\'ach 2, Prague 8, CZ-180 00, Czech Republic
}

\date{Accepted XXX. Received YYY; in original form ZZZ}

\pubyear{2020}

\begin{document}
\label{firstpage}
\pagerange{\pageref{firstpage}--\pageref{lastpage}}
\maketitle

\begin{abstract} 
 Asteroid pairs, two objects that are not gravitationally bound to one another, but share a common origin,
 have been discovered in the Main belt and Hungaria populations. 
 Such pairs are of major interest, as the study of their evolution under a variety of dynamical influences can indicate the time since the pair was created.
 To date, no asteroid pairs have been found in the Jovian Trojans, despite the presence of several
 binaries and collisional families in the population. The search for pairs in the Jovian Trojan population is of particular interest, given the importance of the Trojans as tracers of planetary migration during the Solar system's youth. 
 Here we report a discovery of the first pair, (258656) 2002~ES$_{76}$ and 2013~CC$_{41}$, in the Jovian Trojans.
 The two objects are approximately the same size and are located very close to the L4 Lagrange point. 
 Using numerical integrations, we find that the pair is at least
 $360$~Myr old, though its age could be as high as several Gyrs. The existence of the (258656) 2002~ES$_{76}$--2013~CC$_{41}$ pair implies there could be many such pairs scattered through the
 Trojan population. 
 Our preferred formation mechanism for the newly discovered pair is through the dissociation of an ancient binary system, triggered by a sub-catastrophic impact, but we can not rule out rotation fission of a single object driven
 by YORP torques. A by-product of our work is an up-to-date catalog of Jovian Trojan proper elements, which we have made available for further studies.
\end{abstract}




\section{Introduction}
The discovery of asteroid pairs, two objects sharing a very similar heliocentric
orbit, recently brought yet another piece of evidence into the mosaic of small Solar system
bodies' evolution on short timescales \citep[e.g.,][]{vn08}. Examples of these
couples have been found in the Main belt and Hungaria populations \citep{vn08,pv09,retal11,petal19}.
The similarity between the heliocentric orbits of the two members of an identified asteroid pair
hints at a common and
recent origin for the objects, that most likely involves their gentle
separation from a parent object. Indeed, backward orbital propagation of heliocentric state vectors of the components in many
pairs has allowed researchers to directly investigate the
possibility of their past low-velocity and
small-distance approach \citep[see][for the most outstanding example discovered
so far]{vetal17}. 

The well-documented cases of pairs among asteroids identified to date all feature separation ages
of less than a million years.
\citet{vn08} speculated about three processes that could have led to the formation of those paris:
(i) collisional break-up of a single parent object,
(ii) rotational fission of such an object driven by radiation torques, and 
(iii) instability and separation of the components of a  binary system. 
Whilst each of these possibilities can explain the origin of asteroid pairs, with some being more likely than others for individual pair
cases, evidence has been found that the majority of currently identified pairs were probably
formed through the
rotational fission of their parent object \citep[e.g.,][]{petal10,petal19}.
It is worth noting that Main belt binaries in the same size category (i.e.,
with primary diameters of one
to a few kilometers), are also believed to be primarily formed through the rotational fission of their parent body \citep[e.g.,][]{ph07,maraiv}.
This is an interesting population-scale result that informs us about a leading
dynamical process for few-km size asteroids in the Main belt. It would certainly
be desirable to extend this knowledge to other populations of small Solar system 
bodies.

Attempts to detect orbital pairs in other populations have, to date, either failed or were not
strictly convincing. For instance, the orbital evolution of bodies in the
near-Earth population is very fast and chaotic and, at the same time, the number of
known objects is limited \citep[see, e.g.,][and references therein]{metal19}.
Searches in populations beyond the Main belt were not successful for different
reasons. Whilst dynamical chaos could also be relevant, a more important factor
concerns the smallest size of bodies found at larger distance from the Sun.
The smallest bodies found in Cybele zone, and amongst the Hildas or Jovian Trojans, are
about an order of magnitude larger than the smallest known asteroids in the inner
Main belt or the Hungarias \citep[e.g.][]{emeaiv}. The proposed pair-formation processes
have a characteristic timescale that rapidly increases as a function of parent body size.
For that reason, it is no surprise that, to date, no recently formed ($\leq 1$~My) traditional pairs sharing the same heliocentric orbit have been detected beyond the Main belt.
If any pairs do exist in these distant small-body populations, they should be revealed by their tight configuration in  
proper element space and long-term backward orbital propagation, if the stability in
that particular zone of orbital phase space allows. With that guideline in mind, we focus here on the
Jovian Trojan population. The leap to the Trojan population might appear to contradict
the logical steps of gradually extending our knowledge of Main belt pairs by searches among 
the Cybele or Hilda populations
first. However, we argue that the case of possible Jovian Trojan pairs is actually
more interesting because of that population's entirely different origin.

The Jovian Trojan population consists of two swarms of objects, librating on tadpole trajectories
about the Jovian L4 and L5 Lagrange points.
Indeed, 588 Achilles \citet{Wolf1907588Achillies}
was the first discovered object to serve as an example of a solution to the restricted three body problem
\citep{Lagrange1772}. Whilst originally considered to be just an extension of
the main belt, and particularly the Hilda and Thule populations, towards the orbit of Jupiter, the Jovian Trojans 
were soon realised to be a totally distinct group of objects, with a unique history\citep[see][for a review]{emeaiv}.
Most importantly, the majority of the Jovian Trojans are thought to have
formed in a vast trans-Neptunian disk of planetesimals, at a heliocentric distance beyond $\simeq 20$~au,
and became captured onto their current orbits during the planetesimal-driven instability
of giant planets \citep[see][for review]{Nesvorny2018SSDynam}. The physical properties of the Trojans, such as their material strength or bulk density, are therefore most likely different from most 
of the asteroidal populations, resembling rather those of comets and Centaurs with which they share
the birth-zone. Though relatively stable, the Jovian Trojans can escape their stable region \citep[e.g.][and references therein]{DiSisto2014JupTrojanModels, Holt2020TrojanStability}, and contribute to other populations, most notably the Centaurs \citep[see][and references therein]{DiSisto2019TrojanEscapes}.
An example of this, (1173) Anchises, exhibits significant dynamical instability on timescales of hundreds of millions of years, with the result that it will likely one day escape the Jovian Trojan population and become a Centaur before being ejected from the Solar system, disintegrating, or colliding with one of the planets \citep{Horner2012AnchisesThermDynam}.  

Despite their importance as a source of information on the Solar system's past evolution, fact that the Jovian Trojans are markedly farther from Earth than the Main Belt has made them significantly more challenging targets for study. As a result, our knowledge of the collisional history, binarity, and the presence/absence of pairs in the Trojan population remains far smaller than our knowledge of the main Asteroid belt 
\citep[e.g.][]{MargotAsteroidIVPairsTripBin}. In fact, to date, no confirmed Trojan pairs have been discovered, and the true level of binarity in the population remains to be uncovered.
The most famous confirmed binary in the Trojan population is 
(617) Patroclus, accompanied by a nearly equal size satellite Menoetius \citep[both in the 100~km range; e.g.,][]{metal06,betal15}. The Patroclus-Menoetius system is fully
evolved into a doubly synchronous spin-orbit configuration \citep[see][and references therein]{Davis2020Binary}, and represents an example of the kind of binary systems which
are expected to be common among Trojans. A number of such binaries, comprising two components of almost equal size, have been found amongst the large
trans-Neptunian objects \citep[e.g.,][]{noll20}. This comparison is of particular interest, given that the 
Patroclus system was, in all likelihood, implanted to the Trojan region from the trans-Neptunian region source zone
\citep[e.g.,][]{netal18}. It seems likely that the Patroclus system represents the closest example of an Edgeworth-Kuiper belt binary system. Further information on the Patroclus system will become
available in the in coming decades, as the binary is a target for flyby in 2033
by the \textit{Lucy} spacecraft \citep[e.g.,][]{Levison2017Lucy}. Similar smaller-scale
systems may well exist among the Trojan population
, but their abundance is uncertain. Observationally, such small-scale binaries remain beyond our detection, and theoretical models of their survival depend on a number of unknown parameters \citep[e.g.,][]{netal18,netal20,nv19}.
The existence of Trojan binaries is interesting by itself, but in the context of our work, it is worth noting that, if such binaries exist, they likely serve as a 
feeding cradle for a population of Trojan pairs.

Following this logic, then if the population of pairs among the Trojans
can become known and well characterized, such that their dominant formation process is
understood, that would in turn prove to be a source of new 
information about Trojan binaries. \citet{m93} in his pioneering work on Jovian Trojan orbital architecture noted a
case of L4-swarm objects (1583) Antilochus and (3801) Thrasymedes. Their suspicious
orbital proximity led the author to suggest that they may constitute a genetically
related couple
of bodies. A viable formation process would be through the instability and dissociation of a
former binary (Milani and Farinella, personal communication). Unfortunately, the
Antilochus--Thrasymedes interesting configuration has not since been 
revisited, nor further studied in a more detail.

This background information motivates us to conduct a search for Jovian Trojan pairs.
Unfortunately, even now the problem is not simple, and we consider our work to
be an initial attempt, rather than providing a definitive solution. In section \ref{sel},
we explain our strategy, and describe the difficulties in Trojan pair identification. This strategy 
led us preliminarily identify the Jovian Trojans
(258656) 2002~ES$_{76}$ and 2013~CC$_{41}$ as a potential pair. To test this hypothesis, we attempted to prove that these two bodies could be genetically related using backward
orbital integration, as described in section \ref{int}. In section \ref{disc}, we discuss potential formation processes for the pair, before presenting our 
concluding remarks and a call for observations in section \ref{Concl}. The Appendix~\ref{PropElements} describes
our methods for the construction of Jovian Trojan proper elements. An
up-to-date catalogue of those elements, which we have made publicly  
available online, is actually a fruitful by-product of our work
that may prove useful for future studies. 
We discuss some additional candidate pairs in Appendix~\ref{addpairs}.

\section{Selection of candidate pairs} \label{sel}
The discovery of asteroid pairs was a direct by-product of
a search for very young asteroid families \citep[see][]{netal06,nv06,vn08}. As a
result, the primary ambition was to find pairs that formed recently,
within the last Myr, amongst the Main belt and Hungaria populations. In fact, the
necessity for
proven pairs to be young is essentially related to the method that
allows their identification.

Just like collisional families, asteroid pairs
are identified as a result of the
similarity of their heliocentric orbits. The search for classical collisional families has traditionally been performed
using clustering techniques in proper orbital element space, examining the proper
semi-major axis $a_{\rm P}$, eccentricity $e_{\rm P}$ and the sine of proper inclination
$\sin I_{\rm P}$ \citep[see, e.g.,][for reviews]{bz03,nesaiv}. The use of the proper
elements allows us, with some care, to search for both young and old families. This is
because the proper elements are believed to be stable over much longer timescales
than other types of orbital elements, such as osculating or mean, ideally on a
timescale reaching hundreds of Myrs or Gyrs. 

There are, however, limitations to this method. In the
case of very old families, problems arise from instability of the proper orbital
elements and the incompleteness of the dynamical model used to derive the proper elements. 
A different problem occurs for very young families. The issue has to do with the
huge increase in the number of small-body objects discovered 
over the past decades. Despite the fact
that the very young families and asteroid pairs must have very close values of the
proper orbital elements, it is difficult to statistically discern them from random
fluctuations of background asteroids. Both occur at the same orbital distance in proper element space. 

This fundamental obstacle arises due to the low dimensionality of 
proper element space, which consists of just three independent variables.
In order to separate very young asteroid families and asteroid pairs from the random fluctuations of the background population,
\citet{netal06} and \citet{vn08} realized that this
problem can be overcome if the search is
conducted in a higher-dimensional space. As a result, they used the 
five-dimensional space
of the osculating orbital elements, neglecting just the mean longitude.
The mean orbital elements are also suitable alternative parameters for such an analysis \citep[e.g.,][]{retal11}. In order to
effectively use the two extra dimensions, the searched structures must also be clustered in
secular angles, the longitudes of ascending node and perihelion. This is perfectly justified
for very young families and pairs that are expected to have separated at very low velocities.

Previous searches for these young structures in the space of osculating or mean orbital
elements proved the usefulness of the method, provided the age of the pair was less than about one Myr.
Asteroid pairs will clearly exist that formed
earlier than this limit, but a differential
precession of their secular angles will result in them becoming effectively 
randomized, which will, in turn, render the
identification procedure described above ineffective.
A key point here is that the population of Main belt asteroids is
currently known to very small sizes, with objects detected with diameters of one kilometer, or even smaller. The proposed formation processes for
very young families and pairs are expected to generate
enough pairs within the last Myr
that, even after accounting for discovery biases, we still have some of them in our catalogs.

The situation is, however, different in the case of the Jovian Trojan swarms. First, the
characteristic size of the smallest Trojans is $\simeq5$~km, with few objects being discovered that are
smaller than this limit. Second, the formation processes of putative Trojan pairs, such as a rotational
fission or collisions, are significantly less efficient than in the main belt.
As a result, no identifiable pairs among Trojans are expected to have been
formed in the last $10-30$~Myr, over which time, one would expect secular angles of any such pairs to diverge from each other. 
We conducted a traditional search for pairs in the
five dimensional space of osculating orbital elements
\citep[as in][]{vn08}, but did not find any candidates. If pairs do exist amongst the known Trojans,
their ages must be larger. In that case, however, their secular angles would be randomized,
as is the case for old pairs in the main belt. Our candidate selection method then returns back to the
analysis of the Trojan proper elements, with further considerations based on additional
criteria.
\begin{figure*}
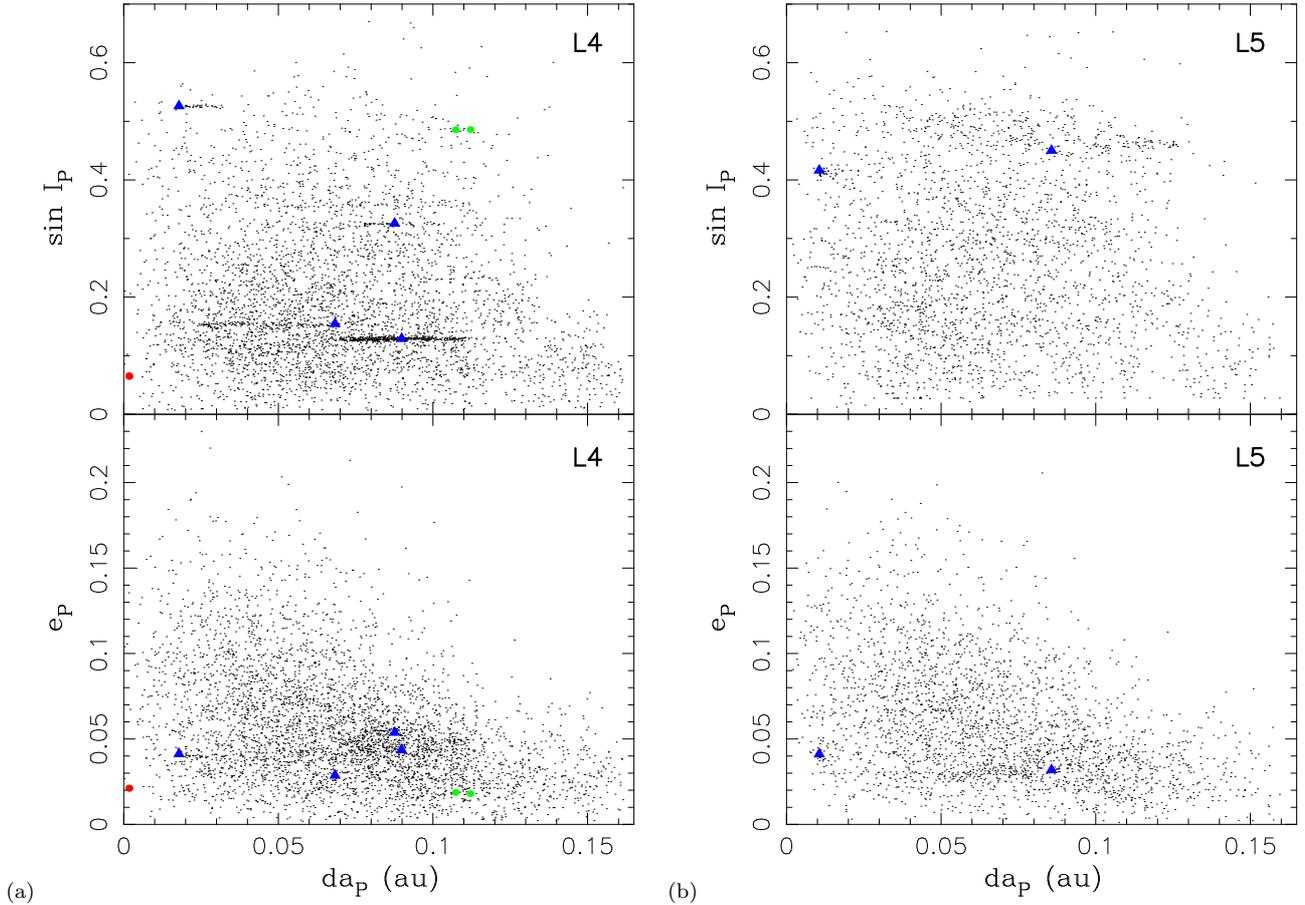

\begin{tabular}{cc}
(a) \includegraphics[width=7.8cm]{f1a.eps}  &
(b) \includegraphics[width=7.8cm]{f1b.eps} \\
\end{tabular}
\caption{Proper orbital elements of Jovian Trojans: semimajor axis $da_{\rm P}$ vs sine
 of inclination $\sin I_{\rm P}$ (top panels), and semimajor axis $da_{\rm P}$ vs
 eccentricity $e_{\rm P}$ (bottom panels). the left panels show the orbits of 4607 objects at
 the L4 libration zone, whilst the right panels show the orbits of 2721 objects at the L5 libration zone.
 These data were computed using the method described in the Appendix~\ref{PropElements} and
 display numbered and multi-opposition orbits as of April~2020. Blue triangles indicate the
 largest objects of the previously identified Trojan families \citep[e.g.,][]{retal16}:
 (a) Eurybates, Hektor, Arkesilaos and 1996~RJ in the L4 zone, and (b) Ennomos
 and 2001~UV$_{209}$ in the L5  zone. The two green circles denote position of Jovian Trojans (1583) and (3801) that were
 previously identified as constituting
 a suspiciously close pair \citep[see][]{m93}. The two
 overlapping red circles denote the location of our proposed pair candidate of (258656) 2002~ES$_{76}$ and
 2013~CC$_{41}$.}
\label{f1}
\end{figure*}

\subsection{A new catalogue of proper orbital elements}
The {\tt AstDyS} website, founded
at the University of Pisa, and currently run by SpaceDys company (see
\url{https://newton.spacedys.com/astdys/}),
is a world
renowned storehouse of proper orbital elements for Solar system minor bodies. It also contains data 
on the Jovian Trojans, namely synthetic proper elements based on mathematical methods
presented in the
pioneering work of \citet{m93}. We also note the work of \citet{br01}, which
discusses an alternative approach to the calculation of Trojan proper elements, but these authors neither
make their results readily available online,
nor update them on a regular basis. For that reason,
one possibility for this study would be to use the {\tt AstDyS} data. However, those data
have at least two drawbacks for our application. First, their last update occurred in June~2017. As a result, they provide
information for a total of 5553 numbered and multi-opposition Jovian Trojans. Given the efficiency of
all-sky surveys, this number has increased significantly in the years since that update, with more than 7000 Jovian Trojans now known for which observations span multiple oppositions.
Second, the proper elements provided at {\tt AstDyS} are given to a precision of just four decimal places
, which is not sufficient for our work. The {\tt AstDyS} database would, as a result, allow the determination of the orbital distance in the proper
element space --Eq.~(\ref{eq-d})-- with only $\simeq 2$ to $5$ m~s$^{-1}$ accuracy, which is insufficient
to characterize the low velocity tail. For both of these reasons, in this work, we decided to determine our own
synthetic proper elements. Details of the approach are given in Appendix~\ref{PropElements}.
Here, we only mention that our proper element definition and mathematical methods follow the
work of \citet{m93}, with substantial differences only for those orbits with very small libration
amplitudes. Previous applications using this technique may be found in \citet{br11} and
\citet{retal16}.

Figure~\ref{f1} shows our results, namely proper elements computed for 7328 Jovian Trojans
(numbered and multi-opposition objects as of April~2020) projected onto the $(da_{\rm P},\sin I_{\rm P})$ and $(da_{\rm P},e_{\rm P})$ planes for the L4 swarm (``Greeks'' leading Jupiter on
its orbit; left panels) and the L5 swarm (``Trojans'' trailing behind Jupiter; right panels).
The L4 swarm is more numerous, partly as a result of four major collisional families that have been recognised in recent years
\citep[e.g.,][]{retal16}, and contains 4607 objects. The smaller L5 swarm contains 
only 2721 known objects, including the 
2001~UV$_{209}$ and Ennomos collisional families. To proceed with an investigation
of the orbital similarity between members of the Trojan population, the basis of the pair and family recognition process, 
one must introduce a metric function in the space of the proper orbital elements. Several choices have been discussed
by \citet{m93}. We opt for the $d_3$ metric, also favoured by the author of that work, though we 
slightly adjust that metric
, such that the orbital distance is given in velocity units. Given two
orbits in the Trojan L4 or L5 proper element space, obviously without mixing the two swarms,
we define their distance $\delta V_{\rm P}$ as a quadratic form using the differences $\delta
a_{\rm P}$, $\delta e_{\rm P}$ and $\delta \sin I_{\rm P}$ as
\begin{equation}
 \delta V_{\rm P} = V_{\rm J} \sqrt{\frac{1}{4}\left(\frac{\delta d a_{\rm P}}{a_{\rm J}}\right)^2
  + 2 \left(\delta e_{\rm P}\right)^2 + 2 \left(\delta \sin I_{\rm P}\right)^2}\; , \label{eq-d}
\end{equation}
where $V_{\rm J}\simeq 13053$ m~s$^{-1}$ and $a_{\rm J}\simeq 5.207$~au are mean orbital velocity
and semimajor axis of Jupiter. \citet{m93} argued that this particular choice of the
coefficients --$(0.25,2,2)$-- helps to equally weight contributions from all three dimensions.
\begin{figure*}
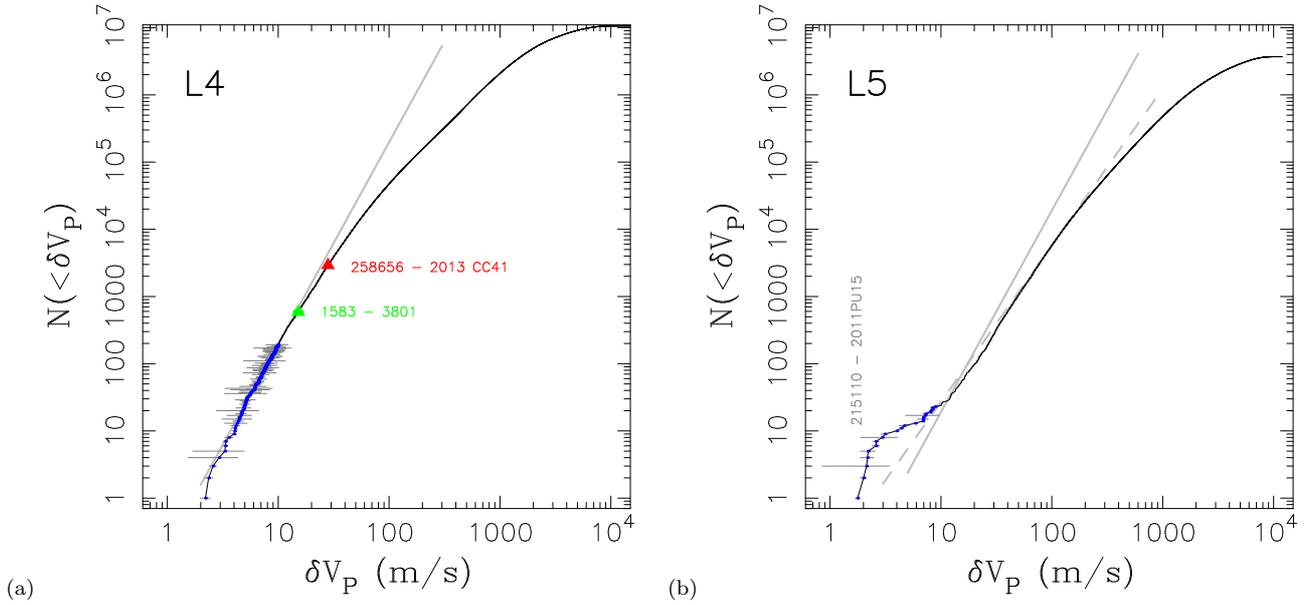

\begin{tabular}{cc}
 (a) \includegraphics[width=7.8cm]{f2a.eps} &
 (b) \includegraphics[width=7.8cm]{f2b.eps} \\
\end{tabular}
\caption{Cumulative distribution $N(<\delta V_{\rm P})$ of Trojans with velocity distance
 $\delta V_{\rm P}$ in the proper elements space using the metric described in Eq. (\ref{eq-d}): the left
 panel presents the results for the 4607 objects of the L4 Trojan swarm, with the right-hand panel showing
 the 2721 members
 of the L5 Trojan swarm. The light gray solid lines indicate the $N(<\delta V_{\rm P})\propto
 (\delta V_{\rm P})^3$ relationship, for reference; curiously, the L5 distribution is better
 matched with $N(<\delta V_{\rm P})\propto (\delta V_{\rm P})^{7/3}$, shown with a dashed gray
 line. Blue symbols denote the population with the smallest $\delta V_{\rm P}$ values, namely
 $\delta V_{\rm P}\leq 10$ m~s$^{-1}$ for both the L4 and L5 Trojans. For sake of interest,
 we also show uncertainties in the determination of $\delta V_{\rm P}$ for these low-velocity
 couples with grey horizontal intervals. Position of three couples of interest is highlighted
 by labels. These are the (1583)-(3801) couple with $\delta V_{\rm P}= (15.2\pm 1.0)$ m~s$^{-1}$ and
 (258656) 2002~ES$_{76}$-2013~CC$_{41}$ couple $\delta V_{\rm P}= (28.2\pm 0.9)$ m~s$^{-1}$ among L4 Trojans, and
 (215110) 1997~NO$_5$-2011~PU$_{15}$ couple with $\delta V_{\rm P}= (1.8\pm 0.1)$ m~s$^{-1}$ among L5 Trojans.}
\label{f2}
\end{figure*}
\begin{figure*}
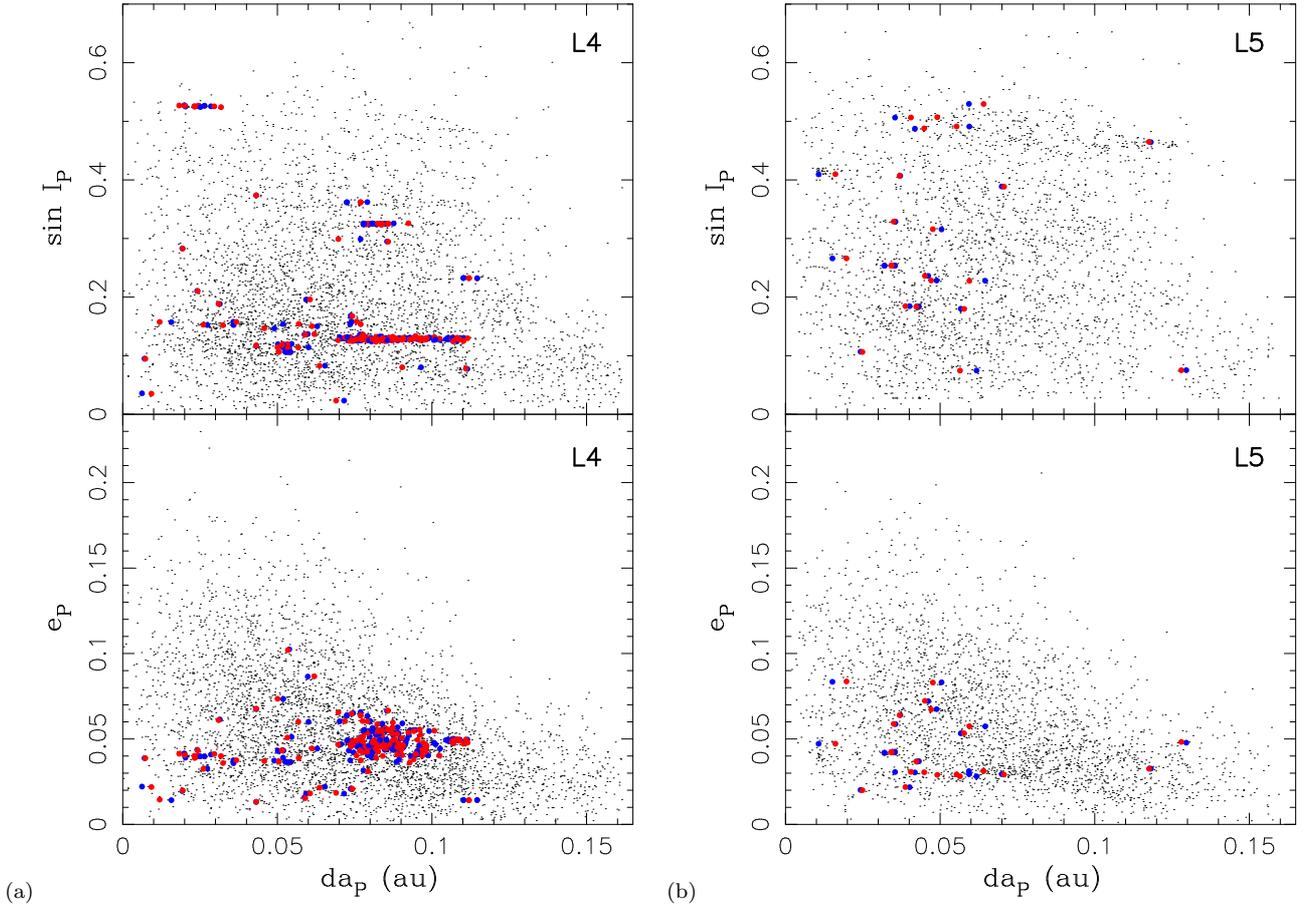

\begin{tabular}{cc}
 (a) \includegraphics[width=7.8cm]{f3a_new.eps}  &
 (b)  \includegraphics[width=7.8cm]{f3b_new.eps}  \\
\end{tabular}
\caption{Proper orbital elements of Jovian Trojans as in Fig.~\ref{f1}. The red
 symbols in both L4 and L5 swarms show couples 
 of Trojans with $\delta V_{\rm P}< 10$
 m~s$^{-1}$, namely the lowest velocity tail in the distributions shown in Fig.~\ref{f2}:
 the primary component of each couple
 is shown using a filled red circle, whilst the secondary is shown using
  a blue circle. In the L4 case their relation to the recognized families is apparent. In
 the L5 case their distribution is more scattered, though some are also associated with
 the 2001~UV$_{209}$ and Ennomos families.}
\label{f3}
\end{figure*}
\begin{figure}
\includegraphics[width=7.8cm]{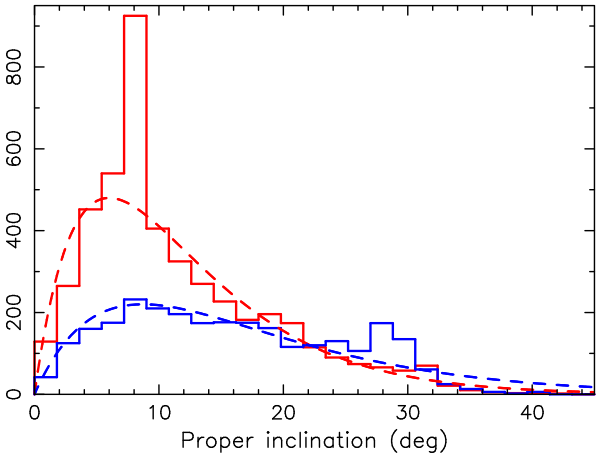}
\caption{Number of Jovian Trojans with proper inclination  $I_{\rm P}$ (in degrees), showing the L4 (red) and L5 (blue) swarms. The dashed lines represent an approximation $I_{\rm P} \exp(-I_{\rm P}/C)$ for the background population (significant peaks due to Trojan families eliminated), where we found $C\simeq 6.0^\circ$ for L4 and $C\simeq 8.7^\circ$ for L5.}
\label{Inc}
\end{figure}

\subsection{Metrics-based analysis}
Given the metric shown in Eq. (\ref{eq-d}), we computed
distances of all possible pairs in the L4 and L5 Trojans swarms, and organized them in the form of
a cumulative distribution $N(<\delta V_{\rm P})$ \citep[see also][for context]{vn08}.
The results of this process are shown in Fig.~\ref{f2}. Whilst the largest $\delta V_{\rm P}$ values of
approximately $V_{\rm J}$ are set by the maximum extension of the stable phase space of tadpole
orbits associated with Jupiter (Fig.~\ref{f1}), the smallest $\delta V_{\rm P}$ values
of order $\sim1-2$ m~s$^{-1}$ are determined by a combination of several factors. The
number of known Jovian Trojans filling the stable orbital space is the first factor, 
compared with the typical smallest values $\delta V_{\rm P}\simeq100$ m~s$^{-1}$ found by
\citet{m93}, who studied just
80 and 94 Trojans in the L4 and L5 swarms, respectively. Additionally, small velocity
differences occur when bodies become organized in structures like families. Last, the
inevitable uncertainty of the proper elements contributes to the
noise in $\delta V_{\rm P}$. We determine the uncertainties of $\delta V_{\rm P}$ by a propagation of the proper element
uncertainties described in the Appendix~\ref{PropElements}. This effect is obviously
not uniform, but organized
in a complicated structure of a chaotic web, generally increasing toward the border of the
stable tadpole zone \citep[see, e.g.,][]{rg06}. Interestingly, the characteristic noise level
from such deterministic chaos is of the order of a few meters per second, about the same as
minimum distances between the orbits, as can be seen in
Fig.~\ref{f2}, where we show uncertainty intervals of $\delta V_{\rm P}$ for the low-velocity tail. 

It is also worth noting that for reasonably
small values of $\delta V_{\rm P}$ (hundred m~s$^{-1}$ or so), one would expect $N(<\delta V_{\rm P})
\propto (\delta V_{\rm P})^3$ provided that: 
(i) Trojans fill the available stable phase space at random, and 
(ii) the weighting coefficients in the metric function (\ref{eq-d}) truly express
isotropy, the exponent 3 is then a measure of the proper element space dimension. For large
$\delta V_{\rm P}$ values the cumulative distributions $N(<\delta V_{\rm P})$ become shallower
because of the finite extent of the stable orbital region. We also note that $N(<\delta
V_{\rm P})$ holds 
global information about the whole L4 and L5 populations, while local
structures, such as families and clusters, are almost not seen in this distribution.

We find it interesting that $N(<\delta V_{\rm P})$ are broadly similar for the L4 and L5 swarms,
but they also differ in some important characteristics, in particular, the smallest and the
largest $\delta V_{\rm P}$ values. This is due to the directly comparable populations of the
two swarms and basically identical volumes of their stable phase space. However, the $\delta
V_{\rm P}< 100$ m~s$^{-1}$ parts of the distributions have a different behaviour when
approximated with a power-law $N(<\delta V_{\rm P}) \propto (\delta V_{\rm P})^\alpha$: (i)
the L4 swarm has the canonical value $\alpha \simeq 3$, while (ii) the L5 swarm is shallower,
with approximately $\alpha\simeq 7/3$. We hypothesize that this difference is caused by a
presence of the prominent Trojan families in the L4 population. Family members efficiently
contribute to the low-$\delta V_{\rm P}$ part of the distribution. Given their small extent,
it is also conceivable that the mutual orbital distribution in families is approximately isotropic.
The L5 population is less influenced by Trojan families, and, as a result, $N(<\delta V_{\rm P})$ may reflect
the parameters of the background Trojan population. This is affected both by the resonances that
sculpt the stable orbital zone in a complicated way and, perhaps
, the initial filling of the
Trojan region by planetesimals. Finally, the weighting coefficients of the metric function
(\ref{eq-d}), that express how differences in semimajor axis, eccentricity and
inclination contribute to the whole, may also slightly affect the result (though our experiments with
small changes in those values
did not yield significant differences). If combined altogether, the $\alpha$ value may be slightly shallower than $3$, such as $7/3$ we found for the L5 population. 

Seeking details that could explain the difference in the population exponents $\alpha$ in further detail, we analysed distributions of the proper elements. The most significant difference concerns proper inclination $I_{\rm P}$. Figure~\ref{Inc} shows L4 and L5 Trojan distributions of $I_{\rm P}$ for all bodies. The
dashed lines are simple approximations with a function $I_{\rm P}  \exp(-I_{\rm P}/C)$, where the adjustable constant $C$ characterizes width of the distribution (the prominent families, such as Eurybates at $\simeq 8^\circ$ among L4 or Ennomos at $\simeq 30^\circ$ among L5, were excluded from the fit). We found $C\simeq 6.0^\circ$ for L4 and $C\simeq 8.7^\circ$ for L5, implying the inclination distribution at L5 is slightly broader. This confirms results in \citet{DiSisto2014JupTrojanModels}
. It is not clear, whether this is due to the details of the capture process, or whether the escapees from the prominent Eurybates and Arkesilaos families in the L4 swarm contribute to the difference, and how it may affect the exponent $\alpha$ of the $N(<\delta V_{\rm P})$ distribution discussed above. A full analysis of these interesting findings is beyond the aims of our work.
Regarding the smallest $\delta V_{\rm P}$ values, neither of the two distribution functions 
$N(<\delta V_{\rm P})$ show a change in behaviour. In the context of our work, this implies 
no hint of a statistically significant population of very close orbits, a tracer of a possible
Trojan pair population. In fact, given the low dimensionality of the proper element space,
this was not unexpected
, given that the asteroid pairs in the Main belt would not manifest themselves using a 
similar analysis. The slight deviation of $N(<\delta V_{\rm P})$ below $\simeq 7$ m~s$^{-1}$ 
velocity to a shallower trend for the L5 swarm is interesting, but likely not 
statistically robust enough to allow firm conclusions to be drawn at the current time. 

We paid some attention to the  smallest-distance couple (215110) 1997~NO$_5$--2011~PU$_{15}$, and could not conclusively prove that it represents
a real pair of related objects (Appendix~\ref{addpairs}). A closer analysis of the second to sixth 
closest couples in the L5 population indicates the  possibility of a very compact cluster 
about Trojan (381148) 2007~GZ$_1$, but its 
status needs to be confirmed with more data in the 
future. In any case, because our interest here focuses on Trojans in the low-velocity 
tail of the $N(<\delta V_{\rm P})$ distribution, seeking putative pairs, we also show 
in Fig.~\ref{f3} location of couples that have $\delta V_{\rm P}< 10$ m~s$^{-1}$ in 
both Trojan swarms. These would be the most logical candidates for further inspection. 

A full frontal approach to this data would be to analyse the results from backward orbital integrations 
for these little more than 200 putative couples using the methods described in
section~\ref{int}. However, this would require a significant
computational effort, and thus we chose to adopt
further criteria for candidate selection. For instance, data in the L4 swarm show that the 
lowest $\delta V_{\rm P}$ couples are strongly concentrated in the recognized families. 
The locally increased density of Trojans in these regions obviously imply small distances 
$\delta V_{\rm P}$, but this also means such couples are most likely not the objects that we seek.
The correlation with Trojan families is somewhat weaker in the L5 swarm, though 
several of the small-distance couples are found in both the Ennomos and 2001~UV$_{209}$ families. 
Other constitute compact clusters scattered in the background population, like that around
(381148) 2007 GZ$_1$, as mentioned above.

Sifting the $\delta V_{\rm P}< 10$ m~s$^{-1}$ couples unrelated to families would 
still leave us with too many candidates to pursue with backward $n$-body simulations. Having experimented with several cases, we 
adopted the strategy of focusing on those low-$\delta V_{\rm P}$ couples characterized by 
(i) the least populated background, and 
(ii) located in the most dynamically stable zones of the orbital phase space. The former condition increases the likelihood
that the candidate couple is a real pair, and not just a fluke, whilst the latter 
condition would allow us to investigate the past orbital configuration of the putative pair across as lengthy a timescale as possible. This is particularly important for pairs in the Jovian Trojan population, since no recently-formed pairs are to be expected, as described above. 
Moreover, the expected large ages of possible Trojan pairs do not allow us to seek their past orbital 
convergence in full six-dimensional Cartesian space of positions and velocities. 
Even the most stable Trojan orbits have an
estimated Lyapunov timescale of about $10-20$~Myr. In this situation, our convergence scheme should rely on the behaviour of secular angles, the longitudes of node and perihelion, and the related 
eccentricity and inclination (section~\ref{int}). It is then advantageous to
suppress the role of the last two elements, the semimajor axis and the mean longitude
, by letting them vary as little as possible
. This favours locations very near the tadpole libration center of either the L4 or L5 swarms, where also the previous
two conditions, low background population and maximum orbital stability, are satisfied.
\begin{figure*}
\begin{tabular}{cc}
 (a) \includegraphics[width=7.8cm]{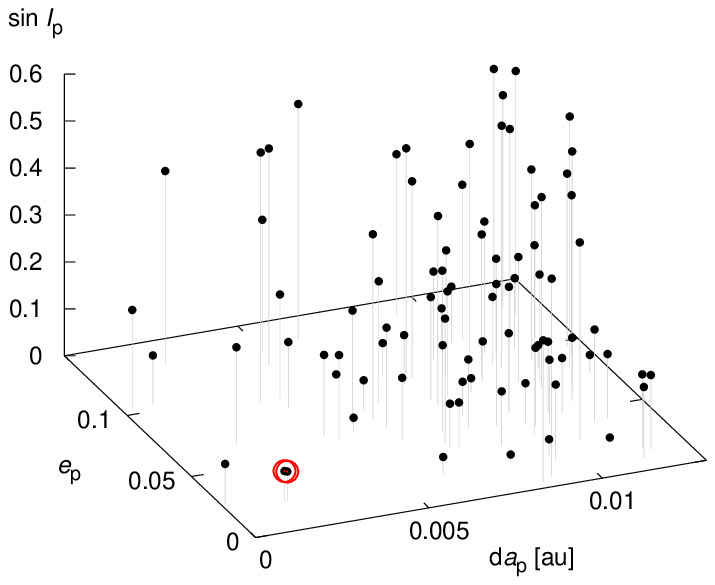} &
 (b) \includegraphics[width=7.8cm]{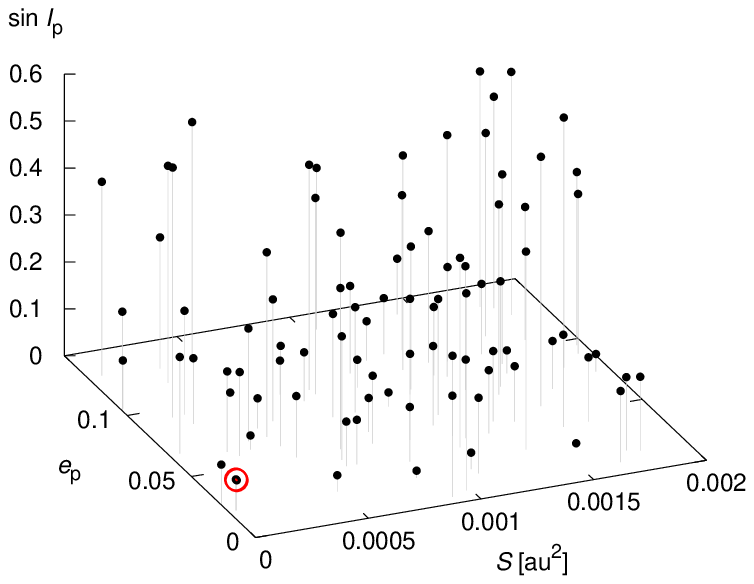} \\
\end{tabular}
\caption{Left (a): The small-libration portion of the L4 stable orbital zone
 in the 3-dimensional proper element coordinates $(da_{\rm P}, e_{\rm P},
 \sin I_{\rm P})$. The proximity to the libration was arbitrarily set by
 $d a_{\rm P} \leq 0.014$~au, whilst the extent of $e_{\rm P}$ and $\sin
 I_{\rm P}$ is limited by orbital stability. We find $91$ objects (black
 symbols) in this zone for our dataset of Trojans. The candidate pair
 (258656) 2002~ES$_{76}$ and 2013~CC$_{41}$ is highlighted with a red circle. The vertical
 intervals help to appreciate 3-dimensional nature of the display. Right (b):
 The same as on the left panel, but the $da_{\rm P}$ was replaced with a
 surface area $S = 4\pi (d a_{\rm P})^2$. In this case, the small-amplitude
 Trojans are distributed more uniformly.}
\label{f4}
\end{figure*}

\subsection{A prospective candidate Trojan pair}
With all these criteria in mind, we found a candidate couple of
L4 objects, (258656) 2002~ES$_{76}$ and 2013~CC$_{41}$. The proximity of these two objects to
the libration center is reflected by the small values of all proper
elements (see Fig.~\ref{f1}), namely $da_{\rm P}\simeq (1.6180\pm 0.0001)\times 10^{-3}$~au, $e_{\rm P}\simeq (2.12713\pm 0.00001)\times 10^{-2}$ and $\sin I_{\rm P}\simeq (6.578\pm 0.003)\times 10^{-2}$ for (258656) 2002~ES$_{76}$, 
and $da_{\rm P}\simeq (1.6890\pm 0.0001) \times 10^{-3}$~au, $e_{\rm P}\simeq (2.10588\pm 0.00001)\times 10^{-2}$ and $\sin I_{\rm P} \simeq (6.427\pm 0.004)\times 10^{-2}$ for 2013~CC$_{41}$. 
The close proximity to L4 also indicates that the pair have been in stable orbits
for the life of the Solar system \citep[e.g.,][]{Holt2020TrojanStability}. For
reference, we also mention their libration amplitude, in the angular
measure, which is only about $0.33^\circ$, resp. $0.34^\circ$. There
are only four other L4 objects in our sample that have smaller libration
amplitudes, and none among the known L5 objects, though these have generally larger
proper eccentricity and/or inclination values. The similarity of the two
orbits is immediately apparent and quantitatively expressed with
$\delta a_{\rm P} \simeq 7.1\times 10^{-5}$~au and $\delta e_{\rm P}
\simeq 2.12\times 10^{-4}$, both with negligible uncertainty, while
$\delta \sin I_{\rm P} \simeq 1.51\times 10^{-3}$ with a small uncertainty
of $4.8\times 10^{-5}$. This uncertainty amounts to about $0.085^\circ$ difference
in the proper inclination. All these values result in the velocity difference
$\delta V_{\rm P} \simeq 28.2\pm 0.9$ m~s$^{-1}$, using our adopted metric
(\ref{eq-d}), dominated by the inclination contribution the contribution from
the difference in proper eccentricities is about $10$\% of the total, and
the difference in proper semimajor axes is negligible). With that said, this
couple would qualify among the closest
in the population if it were not for the slight inclination offset of the two orbits. 

Not much physical information is available about these two objects. Various
databases providing orbital solutions (such as {\tt AstDyS}, {\tt JPL} or
{\tt MPC}) yield an absolute magnitude for
(258656) 2002~ES$_{76}$ in the range $14.0$ to $14.2$
, and values in the range $14.3$ to $14.4$ for 2013~CC$_{41}$. Given the mean
albedo, $p_V\simeq 0.075$, for small Trojans \citep[a value with an admittedly
large scatter; e.g.,][]{getal11,getal12}, we 
estimate their sizes to be $D\simeq 7.0-7.7$~km for (258656) 2002~ES$_{76}$ and $D\simeq 6.4-6.7$~km for 2013~CC$_{41}$.
Unless the assumption of similar albedoes is significantly in error, it is clear that the
two bodies are similar in size, though not exactly the same. No other
physical parameters, such as the rotation period, thermal inertia and/or
spectral colors, are known at the present time.
Further observational follow-up on these objects is therefore highly recommended.

\subsection{Assessment of the statistical significance of the selected pair}
The small libration amplitude zone of the proper element space contains a relatively small number
of bodies, as can be seen in the left panel (a) in Fig.~\ref{f4}. Here, we used the range $d a_{\rm P}
\leq 0.014$~au, expressing the proximity to the libration center, but left $e_{\rm P}
\leq 0.15$ and $\sin I_{\rm P}\leq 0.6$, generally capturing the width of the stable
Trojan phase space \citep{Levison1997JupTrojanEvol, Nesvorny2002Trojans,
Tsiganis2005ChaosJupTrojans, DiSisto2014JupTrojanModels, Holt2020TrojanStability}. We
could have also more strongly restricted the proper eccentricity and inclination values 
, but if this is done too aggressively, it would result in the sample
of observed Trojans available for our analysis becoming too small. With our limits, we find $k=91$ Trojans in the L4 space, including our
candidate pair (258656) 2002~ES$_{76}$ and 2013~CC$_{41}$.

The proper element differences in the (258656) 2002~ES$_{76}$ and 2013~CC$_{41}$ couple are
$\delta a_{\rm P} = 7.11\times 10^{-5}\,{\rm au}$, $\delta e_{\rm P} =
0.000212$, $\delta \sin I_{\rm P} = 0.00151$, much smaller than the scale of the chosen
zone, assuming that all dimensions are taken equally. 
In the first approximation, taking
all dimensions equally, and thus neglecting the weighting coefficients from Eq. \ref{eq-d} which are all of the order of unity,
the $(\delta a_{\rm P},\delta e_{\rm P},\delta \sin I_{\rm P})$ differences in this
couple define a small box of which represents only a $\simeq 1.81\times 10^{-8}$
fraction of the analysed target zone. For statistical calculations
, it is useful to imagine ``numbered'' boxes of the $(\delta a_{\rm P},\delta e_{\rm P},\delta \sin I_{\rm P})$
volume in the whole zone. Their total number of such boxes would then be 
$n\simeq5.53\times 10^7$.

The simplest estimate of the statistical significant of the (258656) 2002~ES$_{76}$-2013~CC$_{41}$ pair 
is based on the assumption that bodies were distributed in the analysed
zone randomly/uniformly. We choose $k$ numbers from $n$ possibilities
(i.e., one for each body from a set of ``numbered'' boxes). Ordered, repeated
selections are given as variations $V'(n,k) = n^k$, while ordered,
non-repeated as $V(n,k) = n!/(n-k)!$. The likelihood that among the
trials the box-numbers do not repeat is simply the ratio $V(n,k)/V'(n,k)$,
and we are interested just in the complementary probability:
\begin{equation}
 p = 1-\frac{V(n,k)}{V'(n,k)} \simeq 7.4\times 10^{-5} \;. \label{like}
\end{equation}
We verified this result by directly running a Monte-Carlo simulation of
the selection process. Thus, we find the probability that the selected
couple is only a random orbital coincidence to be very low. Shrinking
the width of the $e_{\rm P}$ and $\sin I_{\rm P}$ to half the previously
mentioned values did not change our result significantly.

As can be seen in the left panel (a) of Fig.~\ref{f4}, the assumption
of a uniform distribution of background Trojans in the target zone is
fair, but not exactly satisfied. This is the result of the
decreasing number of Trojans towards the libration center (i.e., at very small values $da_{\rm P}$). 
We therefore repeated our  analysis in a different system of coordinates. Keeping $e_{\rm P}$ and $\sin I_{\rm P}$, we now changed
$da_{\rm P}$ with $S = 4\pi(d a_{\rm P})^2$. The background reasoning is
that the libration point, $d a_{\rm P}=0$, represents a center about which
the tadpole orbits move in 3-dimensions. In a Cartesian view centered at
L4 the radial coordinate is to be replaced with the surface area $S = 4\pi(d
a_{\rm P})^2$. Re-mapping and re-binning our analysis in the $(S,e_{\rm P},
\sin I_{\rm P})$ coordinate system, we obtained the situation shown in the right
panel (b) of Fig.~\ref{f4}. Whilst still keeping the same number $k=91$ of
Trojans in the analysed zone, their distribution is now more uniform. Given
the new box-definition by the (258656) 2002~ES$_{76}$ and 2013~CC$_{41}$ couple, we now find
the number of thus defined small boxes to be increased to $n \simeq 2.34\times 10^9$
This is the result of the candidate couple's close proximity to the libration center.
As a result, the likelihood (Eq.~\ref{like}) of the couple being just a fluke in
a uniform distribution of objects now becomes smaller, namely $p\simeq
1.75\times 10^{-5}$. 
\begin{figure}
 \includegraphics[width=8.3cm]{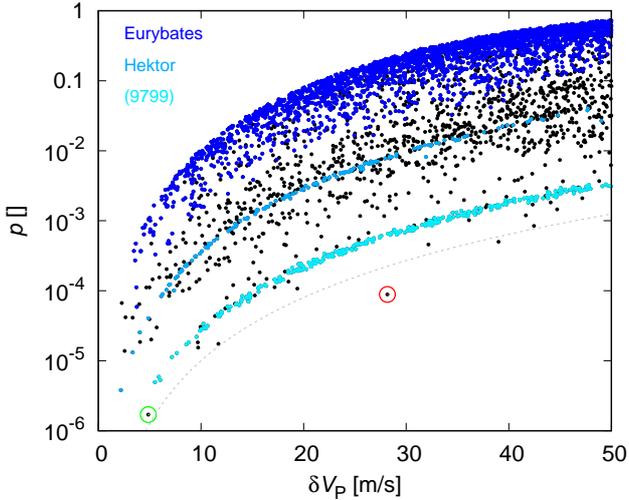}
\caption{The probability~$p$ that a 
pair is random fluke, computed using
 the method described in the text, versus its distance~$\delta V_{\rm P}$,
 computed for all low-velocity pairs in the L4 zone using Eqs.~(\ref{like})
 and (\ref{eq-d}). The pair (258656) 2002~ES$_{76}$ and 2013~CC$_{41}$ is highlighted with a
 red circle. The pair (219902) 2002~EG$_{134}$ and (432271) 2009~SH$_{76}$, discussed briefly in the
 Appendix~\ref{addpairs}, is highlighted with a green circle. The colored
 symbols denote pairs in the identified L4 families: (i) Eurybates (blue),
 (ii) the core of the Hektor family (light blue), and (iii) (9799) 1996~RJ (cyan). 
 The dashed line, $p = 10^{-5}\,(\delta V_{\rm P}/(10\,{\rm m}\,{\rm
 s}^{-1}))^3$, is used to emphasize that the candidate pair (258656) 2002~ES$_{76}$ and
 2013~CC$_{41}$ is an outlier in this population.}
\label{f5}
\end{figure}

The probability $p$, defined and computed for the (258656) 2002~ES$_{76}$--2013~CC$_{41}$ couple above, is appreciably small. It is both interesting and important to compare this result with the similarly defined quantity for other Trojan couples, especially amongst those that have a small $\delta V_{\rm P}$ distance in the
metrics (\ref{eq-d}). This will tell us whether the probability $p$
for (258656) 2002~ES$_{76}$--2013~CC$_{41}$ is sufficiently small in absolute measure for the couple to be considered a true pair, whilst at the same time enabling our algorithm to better connect our $p$ definition with the velocity
metrics used above. Here we analyze the L4-swarm population , but the same approach could equally be applied to the L5 case.

The potentially complicated part of the procedure is that, for each selected couple,
we have to (i) adapt the box size $(\delta a_{\rm P}, \delta e_{\rm P},
\delta\sin I_{\rm P})$, and (ii) the zone size $(\Delta a_{\rm P}, 
\Delta e_{\rm P}, \Delta\sin I_{\rm P})$, as well 
as the position to which the box size refers.
The choice of the latter obviously varies because the local number density of bodies 
differs from place to place. In order to prevent excessively small 
boxes in one of the dimensions (as an example, due to an almost zero
difference $\delta e_{\rm P}\doteq 0$), we use the metric~$\delta V_{\rm P}$ 
as a measure of the ``diagonal'' of the box and we define its respective
volume as $(\delta V_{\rm P})^3/\!\sqrt{3}$. Observing the typical spatial 
variation of the number density of Trojans, we use a fixed value for 
$\Delta a_{\rm P} = 0.02$~au, rejecting pairs with $\delta a_{\rm P} > 
0.3\Delta a_{\rm P}$. In order to prevent a~low number of bodies~$k$ in 
the zone, both $\Delta e_{\rm P}$ and $\Delta\sin I_{\rm P}$ are then 
sequentially increased until $k \ge 50$. Once we set the zone, we again
define its volume as $(\Delta V_{\rm P})^3/\!\sqrt{3}$, with $\Delta V_{\rm P}$
the velocity distance of the corners connected with a diagonal.
The number of boxes~$n$, as well as the probability~$p$, is then computed 
as before (Eq.~(\ref{like})). Obviously, the whole algorithm cannot be
done manually, but an automated computer script was written to run
the method.

The statistical results of our analysis are shown in Fig.~\ref{f5}. The pairs seem to be well organized 
in the $(p,\delta V_{\rm P})$ plane, expressing an overall correlation between
the two quantities. As 
might be expected, the general trend is $p(\delta 
V_{\rm P})\propto (\delta V_{\rm P})^3$, namely volume of the box. Nevertheless, 
the $p$ vs $\delta V_{\rm P}$ values do not follow a single curve, due to 
the local number density being different for each of the couples. Those couples located 
within known families generally have relatively high $p$ values. This is to be 
expected, since  the surrounding zones are densely populated by Trojans,
which causes the dimensions of the zone
to be small. To illustrate this effect,
we colored data for pairs in the largest families in the Fig.~\ref{f5},
identifying those in the
(i) Eurybates family (blue), (ii) the core of the Hektor family (light blue),
and (iii) the (9799) 1996~RJ family (cyan), after \citet{nesaiv}. The Eurybates family, the largest and most populous in the Trojan population, has systematically the largest $p$ values. This is because even a small
zone quickly contains our threshold number of $k=50$ Trojans
. We note that $p\simeq 1$, or even formally larger, just indicates that a couple of Trojans in this zone is fully expected at their distance $\delta V_{\rm P}$.
An exception to this rule
is the (9799) 1996~RJ family, where we find the smallest $p$ values, which are
clearly correlated with $\delta V_{\rm P}$. This is because (9799) 1996~RJ is
a very compact family located in isolation in a high-inclination
portion of the Trojan phase space (see also Fig.~\ref{f1}). For each of the
couples selected in this family, the reference zone needs to be large to
contain the minimum required number of objects.

Whilst the collisional families could clearly
contain dynamical pairs, their recognition is confused
by the locally high background of family members. We therefore
exclude objects located in families
from our work. What remains is then a diffuse background population of Trojans.
For every fixed $\delta V_{\rm P}$ value, there are some background couples for which
$p$ extends to small values. The true Trojan dynamical pairs, namely those objects
genetically related to a common parent, form the basis for our search among this population
of a low-$p$ tail for sufficiently small $\delta V_{\rm P}$ values. There are 
possibly a number of such cases, but amongst them, the one which is the most
outlying from the $p(\delta V_{\rm P})\propto (\delta V_{\rm P})^3$ reference level
shown by the dashed curve in Fig.~\ref{f5} is the case of (258656) 2002~ES$_{76}$--2013~CC$_{41}$ (highlighted with red circle). Its $p$ value is an order of magnitude lower when
compared to couples with similar $\delta V_{\rm P}$ values. This justifies the validity of the (258656) 2002~ES$_{76}$--2013~CC$_{41}$ couple as a true asteroid pair,
based on our statistical analysis alone. There are also some family-unrelated
couples with $p$ values comparable or smaller, and these are briefly discussed
in Appendix~\ref{addpairs}.

In the next section~\ref{int}, we conduct a search for past orbital convergence
of the selected (258656) 2002~ES$_{76}$ and 2013~CC$_{41}$ couple. If successful, this process
add an important piece of evidence justifying the couple as a real
pair of genetically related objects. We explain our methods in detail.
These methods are also briefly applied to several other candidate couples, with less success (Appendix \ref{addpairs}). 

\section{Numerical simulations} \label{int} 
The dynamics of the Jovian Trojans have been extensively studied using both analytical
and numerical means \citep[e.g.,][and references therein]{m93,br01,rg06, DiSisto2014JupTrojanModels, Holt2020TrojanStability}.
Here, we confine ourselves to briefly recalling only the information necessary for
understanding and interpreting our numerical simulations of the (258656) 2002~ES$_{76}$ --
2013~CC$_{41}$ pair. 

As previously noted, the objects in this pair are not
typical, but are instead exceptional representatives of Trojan population.
This is because they reside extremely close to the L$_4$ libration center.
As a result, the evolution of their
semimajor axis $a$ and the resonant argument $\lambda-
\lambda_{\rm J}$ be characterized by  
many small-amplitude and high-frequency terms. Those are, however, of the least importance for our
analysis. More relevant is the behaviour of the eccentricity $e$, the
inclination $I$, the longitude of ascending node $\Omega$, and the longitude of
perihelion $\varpi$. Due to the small values of the eccentricity and inclination
, it is also useful to think about complex non-singular elements $z=e\,\exp(\imath
\varpi)$ and $\zeta=\sin I\, \exp(\imath \Omega)$. In 
linear perturbation theory, a fairly satisfactory zero approximation, both $z$ and $\zeta$ are
represented by a finite number of Fourier terms, namely the proper term
and a few forced planetary terms. A simpler description concerns $\zeta$, whose
Fourier representation is dominated by the proper term with $I_{\rm P}
\simeq 3.7^\circ$, followed only by small contributions from the $s_6$
term, with $I_6\simeq 0.36^\circ$, and a number of significantly
smaller contributions. As a result, the osculating inclination $I$ is well
represented by a constant $I_{\rm P}$ and a periodic term with amplitude
$I_6$. Correspondingly, the osculating longitude of the ascending node, $\Omega$, steadily
circulates with a period given by the proper $s$ frequency, and experiences only
very small perturbation from the $s_6$ term. The evolution of $z$ is more
complicated because it is represented by three terms of comparable
amplitude. The largest-amplitude contribution, $\simeq0.044$, is provided by the
term with frequency $g_5$, followed by proper $g$ and $g_6$ terms with
comparable amplitudes of $\simeq0.021$ and $\simeq0.015$. Whilst still
very simple in the Cartesian representation of $z$, the polar variables
in this plane (i.e. the eccentricity and especially longitude of
perihelion) exhibit a non-linear evolution, characteristic of many
low-eccentricity asteroid orbits.

\begin{figure}
 \includegraphics[width=7.8cm]{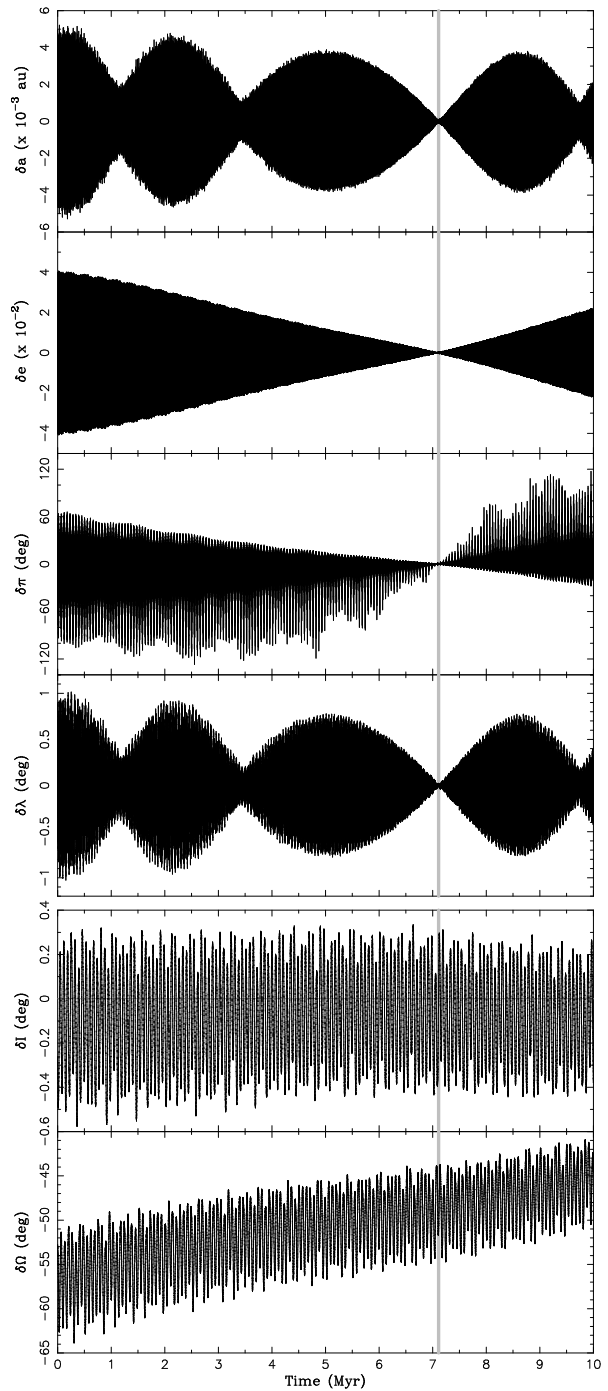}
\caption{Differences between the osculating orbital elements of
 (258656) 2002~ES$_{76}$ and 2013~CC$_{41}$ from a $10$~Myr backward integration
 of their nominal orbits. Gravitational perturbations from all
 planets were included and an invariable-plane reference system
 used. The differences of semimajor axis $\delta a$, eccentricity
 $\delta e$, longitude of pericenter $\delta \varpi$ and 
 longitude in orbit $\delta \lambda$ (top four panels) indicate a
 simultaneous collapse to near zero values at $\simeq7.11$~Ma
 (gray vertical line). In contrast, the differences of inclination
 $\delta I$ and longitude of ascending node $\delta \Omega$ (the bottom two
 panels) do not converge at that epoch: the nodal longitudes of
 the two objects are still $\simeq50^\circ$ away from each
 other, and the inclination difference shows steady oscillation
 about the mean value of $\simeq-0.08^\circ$, namely a difference
 in their proper inclinations. The steady trend in $\delta \Omega$
 has a slope $\simeq0.004$ arcsec~yr$^{-1}$, very close to the
 difference in proper frequencies $s$ of (258656) 2002~ES$_{76}$ and 2013~CC$_{41}$.}
\label{f6}
\end{figure}
\subsection{Short-term simulations}
Equipped with this knowledge, we can now turn to investigating
the common origin of (258656) 2002~ES$_{76}$ and 2013~CC$_{41}$. In studies of asteroid
pairs, researchers seek to demonstrate a convergence of 
heliocentric orbits of the proposed pair at some moment in the past \citep[e.g.,][]{vn08}.
This is considered to be the origin of the two objects from a common
parent body, and the corresponding time in the past representative of the
age of the pair. As typically achievable ages of the asteroid pairs
in the Main belt are less then one Myr, with many less than
$100$~kyr, a convergence is often sought in Cartesian space. This
approach means to demonstrate that the two orbits meet at the same
point in space and have a very small relative velocity. 

The same condition can be expressed in heliocentric orbital elements by
making them basically equal at the formation moment of the pair.
For this work, we find it markedly more useful to work
with the orbital elements of our candidate pair,
as they can teach us more readily about the evolution of the
orbits of the two objects. Therefore, in Fig.~\ref{f6}, we show the results of our
initial numerical experiment. We provide the differences between
the osculating heliocentric elements of the nominal orbits of (258656) 2002~ES$_{76}$ and 2013~CC$_{41}$
over a short time interval of the past $10$~Myr. We use the {\tt
swift\_rmvs4} integrator \citep{ld94} which allows us to efficiently
include gravitational perturbations from all eight planets. The integration timstep used
was $3$~days, and the state vectors of all propagated bodies, planets and
the two Trojans, were output every $50$~years. We use a reference
system defined by the invariable plane of the planetary system. The
initial conditions of (258656) 2002~ES$_{76}$ and 2013~CC$_{41}$ at MJD58800 epoch
were obtained from the {\tt AstDyS} website.

The differences in the orbital elements shown in Fig.~\ref{f6} oscillate
with the dominant frequencies identified by the analysis of $z$ and
$\zeta$ themselves. For instance, the principal periodicity seen in
$\delta I$ and $\delta \Omega$ corresponds to the frequency $s_6-s$,
whilst the principal periodicity seen in $\delta e$ and $\delta \varpi$
corresponds to frequencies $g$ and $g_5-g_6$. Differences $\delta a$
and $\delta \lambda$ are characterized by higher frequencies, such as
the planetary orbital frequencies, the libration frequency, and then
followed by a ``forest'' of lower frequencies starting with $g$.

We also note a markedly different behavior of $\delta \varpi$ and $\delta
\Omega$, which can be understood from the above mentioned description
of the $z$ and $\zeta$ non-singular elements of the two objects.
Observing the general behavior of the amplitude in the $(\delta a, \delta
e,\delta \varpi,\delta \lambda)$ terms, we note a curious fact that those amplitudes become
very small simultaneously for semimajor axis, eccentricity,
longitude of perihelion and longitude in orbit $\simeq 7.11$~Myr ago
(upper four panels in Fig.~\ref{f6}). However, any
hope for a clear orbital convergence at
that epoch is removed by looking at behavior of the inclination and
longitude of ascending node differences (bottom two panels in Fig.~\ref{f6}). We note that $\delta
I$ keeps steadily oscillating about a mean value of $\simeq -0.08^\circ$,
namely a difference in the proper inclinations of (258656) 2002~ES$_{76}$ and
2013~CC$_{41}$, without the amplitude of those oscillations showing any tendency to shrink.
At the same time, the nodal difference stays large, and only slowly
decreases from $\simeq -56^\circ$ to $\simeq -45^\circ$. This rate of
decrease in $\delta \Omega$ fits perfectly the difference in proper
frequencies $s$ of the two objects as to be expected. Hence some $\simeq7.11$~Myr ago, 
the two orbits had basically identical $(a,e,\varpi,
\lambda)$ values, but the nodes were still offset by about $50^\circ$.
This is inconsistent with any believable low-velocity separation of
the two objects from a common parent body at their origin. Whilst
inconclusive about the origin of the (258656) 2002~ES$_{76}$ and 2013~CC$_{41}$ couple,
this $10$~Myr integration provides 
useful hints for further analyses.

\begin{figure*}
 \includegraphics[width=15.5cm]{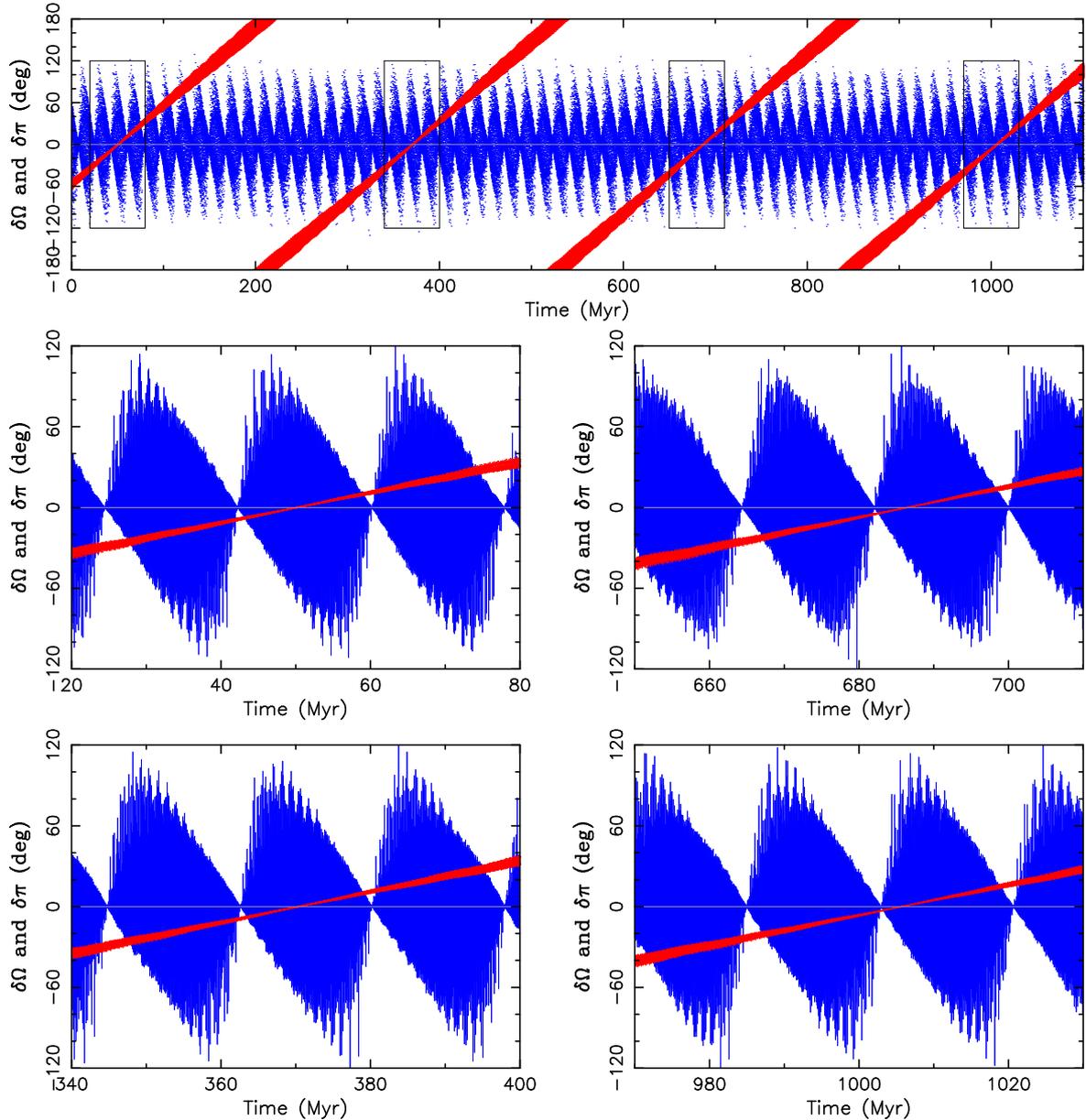}
\caption{the long-term behavior of the difference in osculating nodal
 and perihelion longitudes $\delta\Omega$ (red) and $\delta\varpi$
 (blue) for the nominal orbits of (258656) 2002~ES$_{76}$ and 2013~CC$_{41}$. The top
 panel shows the results from a backward integration in time to
 $1200$~Myr. The four panels below show a zoom around the
 configurations where $\delta\Omega$ becomes small, also
 indicated by the black rectangles in the top panel. As inferred
 from data in Fig.~\ref{f6}, the first such situation occurs
 $\simeq 50$~Myr in the past, and repeats with a period of $\simeq
 320$~Myr. The configuration of the nominal orbit becomes closest
 to true convergence at $\simeq 680$~Myr and $\simeq 1010$~Myr in
 the past (right middle and bottom panels).}
\label{f7}
\end{figure*}
\subsection{Long-term simulations}
Extrapolating the trend seen in Fig.~\ref{f6}, we can
estimate that the nodes of (258656) 2002~ES$_{76}$ and 2013~CC$_{41}$ became coincident some $50$~Myr
ago. Obviously, this is only the first such configuration in the
historical evolution of the two objects. Assuming orbital stability, we 
also predict that the configuration will
repeat with a $\simeq 320$~Myr periodicity. To probe the long-term
changes in the orbital architecture of the (258656) 2002~ES$_{76}$ -- 2013~CC$_{41}$ couple,
we extended our previous simulation to $1200$~Myr in the past. We
note in passing that the necessity to seek this pair's age over
such a long timespan forces us to abandon any hopes of finding
a convergence in Cartesian coordinates. This is because of the small but non-negligible
chaoticity of the integrated orbits, and principally results from an
uncertainty in the thermal accelerations that the objects would experience (as discussed below). Both of these factors would
require a large number of clones of (258656) 2002~ES$_{76}$ and 2013~CC$_{41}$ to investigate their past
histories, and thus are computationally prohibitive to pursue. We therefore choose to downsize the dimensionality of the
space where a convergence is quantified, and focus on the behavior
of secular evolution in just the non-singular elements $z$ and $\zeta$.
Figure~\ref{f7} shows the differences between the osculating $\delta\Omega$ and
$\delta\varpi$ of the two objects, and pays special attention to the time interval
near $\delta\Omega\simeq0$ configurations. 

As expected, the first such configuration occurred about $50$~Myr ago. However, a closer
look at the relevant panel of Fig.~\ref{f7} indicates that 
suitable orbital convergence conditions did not occur at that time. 
Unlike $\simeq7.11$~Mya, the orbital planes converge, but the perihelion longitudes
are at the maximum of their oscillations. An even closer look at the
epochs near nodal convergence shows that when $\delta\varpi$ crosses
zero, $\delta e$ is large, and vice versa. Once again, we therefore find that the
conditions of a low-velocity separation of the two orbits cannot be met at that epoch.

Inspecting further epochs of nodal crossing, as shown in 
Fig.~\ref{f7}, we conclude that $\delta\Omega\simeq 0$ in fact never exactly coincides
with $\delta\varpi \simeq0$, a convergence pre-requisite. Here,
however, we must revisit some of the assumptions made in our
simulation. In particular, recall that (i) we used 
only nominal realizations of the orbits of both (258656) 2002~ES$_{76}$ and 2013~CC$_{41}$, and
(ii)we included only gravitational perturbations from planets in our
dynamical model. Both of these approximations are insufficient for a full analysis of our
problem (see a similar discussion of the attempts to determine the origin
of young asteroid clusters/families and pairs in \citet{nv06},
or \citet{vn08}).

First, the nominal orbital solution represents the best-fit of the
available astrometric data. The inevitable uncertainties of the latter
implies the uncertainty of the orbital fit itself. Well-behaved
orbital solutions are represented by fixed confidence-level regions
in the six-dimensional orbital space, using an ellipsoidal geometry,
mathematically expressed by elements organized in the covariance
matrix. Each orbit starting in a high confidence-level zone ($\geq
80-90$\%, say) is statistically equivalent to the best-fit solution.
whilst initially very compact, these different solutions typically
diverge with time. We thus need to consider in our simulation not only
the best-fit orbits, but also a sample of those starting from the 
high-confidence zone. We call these ``geometrical clones''.

The second issue that needs to be considered is the validity of the dynamical 
model used. the long-term dynamics of small objects are known to be subject to perturbations 
due to the thermal acceleration known as the Yarkovsky effect 
\citep[e.g.,][]{betal06,vokaiv}. Nominally, within the Trojan population, 
objects are only minimally affected by the Yarkovsky effect \citep{wh17, hetal19}, 
which has the greatest influence at
smaller sizes. However, the two components in the (258656) 2002~ES$_{76}$-2013~CC$_{41}$ 
couple are well within this size range, and so it is warranted to see what dynamical effects might be
produced by Yarkovsky accelerations. Since none of the parameters needed for
evaluation of the thermal accelerations, such as the rotation state, the surface thermal inertia, and the bulk density, are known for either (258656) 2002~ES$_{76}$ or 2013~CC$_{41}$, we 
need to consider a suite of potential orbit histories, each generated by numerical integration 
of test particles experiencing the a range of physically plausible thermal accelerations
. These will be called the Yarkovsky clones. We also note that the 
effect of thermal accelerations was included in
{\tt swift\_rmvs4} using the same method as described in \citet{vn08}.

\subsection{Clone sets}
We conducted two sets of numerical simulations, one considering only the geometrical 
clones (section~\ref{geom}), and the other considering only the Yarkovsky clones (section~\ref{yarko}) of (258656) 2002~ES$_{76}$ and 2013~CC$_{41}$. 
In each simulation set, we include the nominal orbit of the objects, complemented by
a set of 20 clones. We ran a backward integration of all orbits for $1.5$~Gyr with an integration
timestep of $3$~days. 
Every $500$~years, we evaluated the differences 
between the osculating orbital elements of the 
21 realizations of (258656) 2002~ES$_{76}$ with each of those of 2013~CC$_{41}$, and searched for 
the 
possibility of a convergent configuration. To quantify the latter, we used two 
conditions. First, as in \citet{nv06}, we evaluated the target function
\begin{equation}
 \delta V = na \sqrt{(\sin I\,\delta\Omega)^2 + 0.5\, (e\,\delta\varpi)^2}
 \; , \label{eq1}
\end{equation}
where $(n,a,e,I)$ are the arithmetically-mean values of the mean motion, semimajor 
axis, eccentricity and inclination of the two considered clones, and $\delta\Omega$ 
and $\delta\varpi$ are the differences between
the osculating longitude of the ascending node and 
perihelion for the two clones, respectively. This way, $\delta V$ has the dimension of velocity, and is constructed 
to provide, in a statistically mean sense, the necessary velocity perturbation required 
for a transfer between the secular angles of the two orbits. However, the analysis
of the results presented in Fig.~\ref{f7} has shown that even a configuration with potential $\delta\Omega 
\simeq 0$ and $\delta\varpi\simeq 0$, and therefore $\delta V\simeq 0$,
is not enough to guarantee
a satisfactory orbital convergence, provided that $\delta e$ and 
$\delta I$ are simultaneously large. For that reason, we admit as a potentially 
convergent configuration a case where the orbits of the two clones satisfy
\begin{itemize}
\item $\delta V\leq V_{\rm lim}$, where $V_{\rm lim}$ is some small value, we
      use typically $1-3$ m~s$^{-1}$, and
\item $\delta e\leq e_{\rm lim}$ and $\delta I\leq I_{\rm lim}$, where again we use
      suitably small values of $e_{\rm lim}\simeq 5\times 10^{-4}$ and $I_{\rm lim}
      \simeq 0.1^\circ$ namely differences in the corresponding proper elements of
      (258656) 2002~ES$_{76}$ and 2013~CC$_{41}$.
\end{itemize}
We output information about these potentially converging configurations for further 
analysis. In the next two sections, we comment on the results of 
our numerical experiments that use geometrical (section \ref{geom}) and Yarkovsky clones (section \ref{yarko}) separately.
\begin{figure}
 \includegraphics[width=\columnwidth]{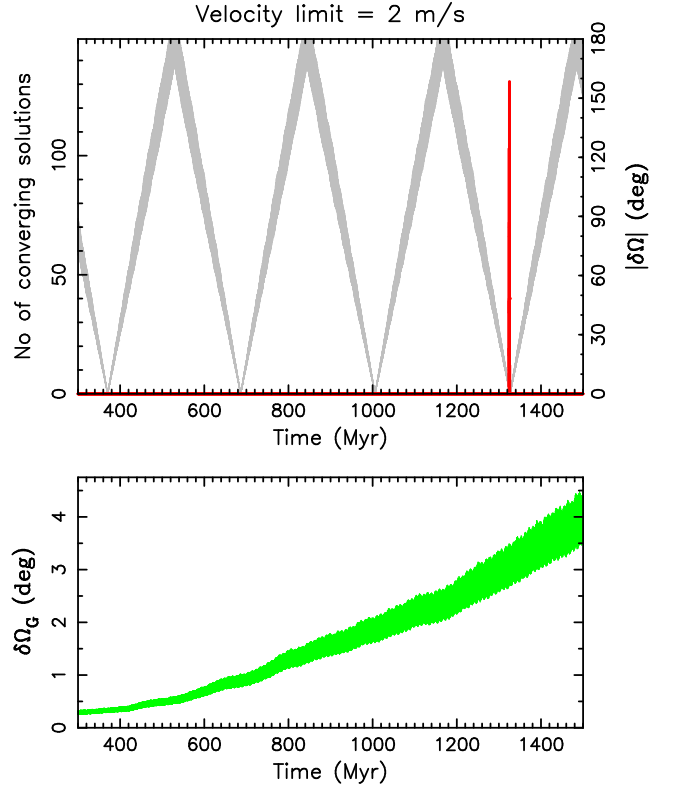}
\caption{The statistical distribution of convergent solutions for geometric clones of (258656) 2002~ES$_{76}$
 and 2013~CC$_{41}$ from simulations of the nominal orbits of the two objects, plus 20 clones
 each, using the velocity cutoff $\delta V\leq 2$ m~s$^{-1}$, and
 eccentricity and inclination limits discussed in the text. Abscissa
 is time to the past starting from $300$~Mya (there are no earlier
 solutions).The left ordinate in the upper two panels gives the number
 of recorded solutions in $50$~kyr bins (red histogram). The gray
 line gives $|\delta \Omega|$ of the nominal orbits of (258656) 2002~ES$_{76}$'s and 2013~CC$_{41}$ 
 (see the right ordinate and the red line on
 Fig.~\ref{f7}), aiming to aid interpretation of the results. The
 green line in the bottom panel shows the maximum difference in
 the longitude of the ascending node between the clones of (258656) 2002~ES$_{76}$ and the longitude
 of ascending node of its nominal orbit; up to about $200$~Myr this trend is
 nearly linear, but becomes more complicated beyond this epoch due
 to very weak orbital chaos.}
\label{f8}
\end{figure}

\subsubsection{Geometric clones} \label{geom} 
Information about the orbit determination, needed for a construction of the 
geometrical clones, was taken from the {\tt AstDyS} database. the orbits of both 
(258656) 2002~ES$_{76}$ and 2013~CC$_{41}$ are rather well constrained, reflecting numerous 
astrometric observations. Even the poorer of the two, 2013~CC$_{41}$, was 
observed over seven oppositions, leading to a fractional accuracy of $\simeq
10^{-7}$ in the semimajor axis, $a$, and the Cartesian components of the 
non-singular elements, $z$ and $\zeta$.
Only the mean longitude, $\lambda$, has a slightly worse accuracy, namely $\simeq 2\times 10^{-5}$ 
degrees. These are the characteristic differences between the six orbital 
osculating elements ${\bf E} = (a,z,\zeta,\lambda)$ of the clones in 
$\simeq 68$\% confidence zone and the best-fit solution ${\bf E}_\star$. 
The solution is given at the initial epoch MJD58800. Complete information 
about the parameters of the six-dimensional confidence zone ellipsoid in the 
space of elements ${\bf E}$ is given by the covariance and normal matrices, 
also provided at the {\tt AstDyS} website. Denoting $\mathbf{\Sigma}$ the 
normal matrix, we may construct the initial orbital elements ${\bf E}$ of 
the geometric clones using
\begin{equation}
 {\bf E} = {\bf T}^{\rm T} \mathbf{\xi} + {\bf E}_\star\; , \label{e2}
\end{equation}
where $\mathbf{\xi}$ is a six-dimensional vector whose components are random 
deviates of normal distribution (with variance equal to unity), and the 
matrix ${\bf T}$ satisfies ${\bf T}^{\rm T}{\bf T} = \mathbf{\Sigma}$ 
\citep[e.g.,][]{mgbook}; ${\bf T}$ is obtained using the Cholesky 
decomposition method. As mentioned above, we constructed 20 geometric 
clones of both (258656) 2002~ES$_{76}$ and 2013~CC$_{41}$ at the initial epoch of our simulation.

The bottom panel of Fig.~\ref{f8} shows the maximum nodal difference 
between the clones of (258656) 2002~ES$_{76}$ and its nominal orbit. Tiny differences 
between the orbital parameters imply that the $s$ frequency of the clone 
orbits is not exactly the same as that of the nominal orbit. However, 
the stability of this orbital zone ensures that the configuration of the 
clone orbits does not evolve, and thus initially the nodal divergence 
is basically linear in time. Only beyond about $0.5$~Gyr does the divergence 
become stronger than linear. This is an expression of a very weak 
instability that manifests itself in the behavior of the secular angle 
solely Gyr timescales 
. The formal Lyapunov timescale of the orbits of both (258656) 2002~ES$_{76}$ 
and 2013~CC$_{41}$ is only $\simeq 20$~Myr (see the {\tt AstDyS} 
database). This implies that a divergence in $\lambda$ is dominant, 
whilst the divergence in the secular angles is slower, as shown in Fig.~\ref{f8}. At 
$1$~Gyr, the nodal longitudes of clones of (258656) 2002~ES$_{76}$ are thus spread over a 
$\simeq 2^\circ$ range. A similar, and potentially slightly
larger, effect is seen among the clones of 2013~CC$_{41}$, principally due to their 
larger differences at the initial epoch. This divergence may overcome the
difficulties we experienced in attempting to find an epoch at which the nominal orbits 
achieve a converging configuration. 
For instance, in the bottom right panel of Fig.~\ref{f7}, we note 
that the nodal difference of the nominal orbits misses the epoch at 
which the difference of pericenters basically shrinks to zero by about 
$3^\circ$ at $\simeq 1$~Gyr. This may be compensated for if the orbits of
suitable clones are used, instead of the nominal orbits. Obviously, a 
satisfactorily large nodal spread of the clone orbits must be attained.

The top panel of Fig.~\ref{f8} shows the statistical distribution of the
converging geometric clones of the two Trojans, organized in $50$~kyr wide bins. 
Obviously, the rather small number of clones in our test run does not
allow us to probe the convergence properties in 
great detail. For that reason, and with the rather tight limit $\delta V\leq 2$ m~s$^{-1}$
chosen, the possible solutions cluster only near the $\simeq 1325$~Myr
epoch, though we note that, if a looser criterion $\delta V\leq 4$ m~s$^{-1}$ was chosen,
more solutions would also exist at $\simeq 1003$~Myr. Taken naively
at a face value, we would conclude a possible origin of the (258656) 2002~ES$_{76}$
-- 2013~CC$_{41}$ couple at this time in the past, if the couple are not older than
$1.5$~Gyr, beyond which we did not continue our simulation. However,
as is often in the case of a pair configuration which is not very young,
the so far neglected thermal accelerations in the dynamical model can prove to be a source of considerable uncertainty.This is analysed in section \ref{yarko}.

\subsubsection{Yarkovsky clones} \label{yarko} 
Our Yarkovsky clones all have the same initial conditions as the 
nominal orbit, but they differ in the magnitude of thermal
accelerations used for their orbital 
propagation. As in \citet{vn08}, 
we approximate thermal accelerations using a simple transverse
component with the magnitude inversely proportional to the square
of the heliocentric distance. The magnitude of this acceleration
is adjusted such that the resulting change in the semimajor axis
$da/dt$ matches predictions from the theoretical formulation of Yarkovsky effect 
\citep[see also][where a classical formalism used in cometary
dynamics was adopted]{fetal13}. In order to estimate plausible
$da/dt$ values, we use a simple approach describing the diurnal
Yarkovsky effect for a spherical body on a circular heliocentric
orbit, presented in \citet{v98}. We use the following set of
physical parameters: the surface thermal conductivity $K\simeq
0.01-0.03$ W~m$^{-1}$~K$^{-1}$, the surface thermal inertia
$\Gamma\simeq 100-200$ [SI units] \citep[for both see][]{delaiv},
the bulk density $\rho\simeq 1.5$ g~cm$^{-3}$ \citep[e.g.,][]{c12},
rotation period $P\simeq 100-500$~hr, and size $D\simeq 7$~km.
The maximum semimajor axis drift rate at zero obliquity is then
$(da/dt)_{\rm max}\simeq (0.15\pm 0.07)\times 10^{-4}$ au~Myr$^{-1}$.
Our choice of a slow rotation period is tied to the working
assumption that (258656) 2002~ES$_{76}$ and 2013~CC$_{41}$ are indeed a real Trojan
pair. We argue in section~\ref{dissBinary} that the most plausible formation mechanism for such a pair is the 
destabilization of a Trojan binary. If this is indeed the case, then
before their separation, the two components were most likely spin-orbit synchronized to
periods of $\geq 100$~hr \citep[e.g.,][]{netal20}. If, however, the
formation mechanism of the pair was different, such as the YORP-driven
fission of a parent object (see section \ref{rotFission}), the rotation periods $P$ of (258656) 2002~ES$_{76}$ and
2013~CC$_{41}$ could well be as short as a few hours.
In that case, $(da/dt)_{\rm max}$ would be smaller by a factor of $3$ to $5$. Indeed,
as a confirmation of our reasoning, we note that scaling 
the value of the detected Yarkovsky signal $19\times 10^{-4}$ au~Myr$^{-1}$ for the $500$~m
size near-Earth asteroid 101955 Bennu with $P\simeq 4.3$~hr
\citep[e.g.,][]{cetal14}, we would have $(da/dt)_{\rm max}\simeq
0.06 \times 10^{-4}$ au~Myr$^{-1}$. In our simulation, we consider
only the case of long rotation periods, and fix $(da/dt)_{\rm max}\simeq
0.15\times 10^{-4}$ au~Myr$^{-1}$. For each of the two Trojans,
(258656) 2002~ES$_{76}$ and 2013~CC$_{41}$, we consider the nominal orbit with
$da/dt=0$, and $20$ Yarkovsky clones. In both cases, $10$ clones
have positive $da/dt$ and $10$ clones have negative $da/dt$.
Additionally, because in the case of the diurnal variant of the
Yarkovsky effect $da/dt\propto \cos\gamma$, where $\gamma$ is the
spin axis obliquity, the positive/negative close $da/dt$ values
uniformly sample the interval $0$ to $(da/dt)_{\rm max}$, resp.
$-(da/dt)_{\rm max}$ to $0$.
\begin{figure}
 \includegraphics[width=\columnwidth]{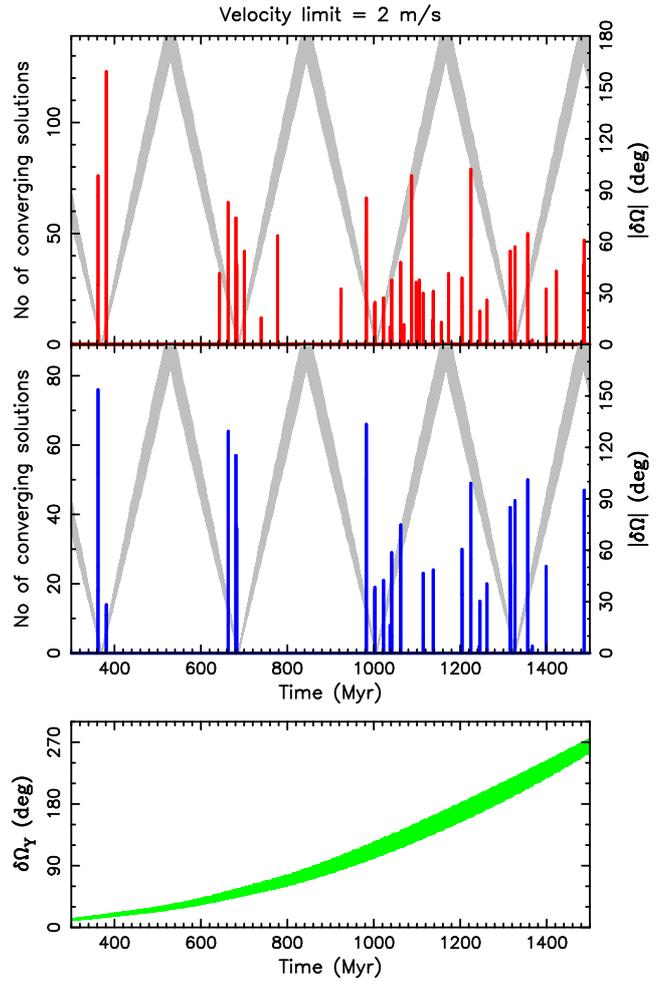}
\caption{the statistical distribution of convergent solutions for the Yarkovsky clones (nominal orbits plus 20 clones each) of (258656) 2002~ES$_{76}$
 and 2013~CC$_{41}$, using the velocity cutoff $\delta V\leq 2$ m~s$^{-1}$, and
 eccentricity and inclination limits discussed in the text. Abscissa
 is time to the past starting from $300$~Mya (there are no earlier
 solutions). the left ordinate in the upper two panels gives the number of
 recorded solutions in $50$~kyr bins. The top panel (red histogram)
 gives the number of solutions for all possible combinations of clones.
 The middle panel (blue histogram) for the case when only clones with
 the same sign of $da/dt$ were compared. The gray line gives $|\delta
 \Omega|$ of the (258656) 2002~ES$_{76}$'s and 2013~CC$_{41}$'s nominal orbits (see the
 right ordinate and the red line on Fig.~\ref{f7}), aiming to aid
 interpretation of the results. The green line in 
 the bottom panel shows
 the difference in the longitude of ascending node between the Yarkovsky clone with
 maximum positive drift rate $(da/dt)_{\rm max}$ and the nominal orbit
 of (258656) 2002~ES$_{76}$.}
\label{f9}
\end{figure}
\begin{figure*}
 \includegraphics[width=15.7cm]{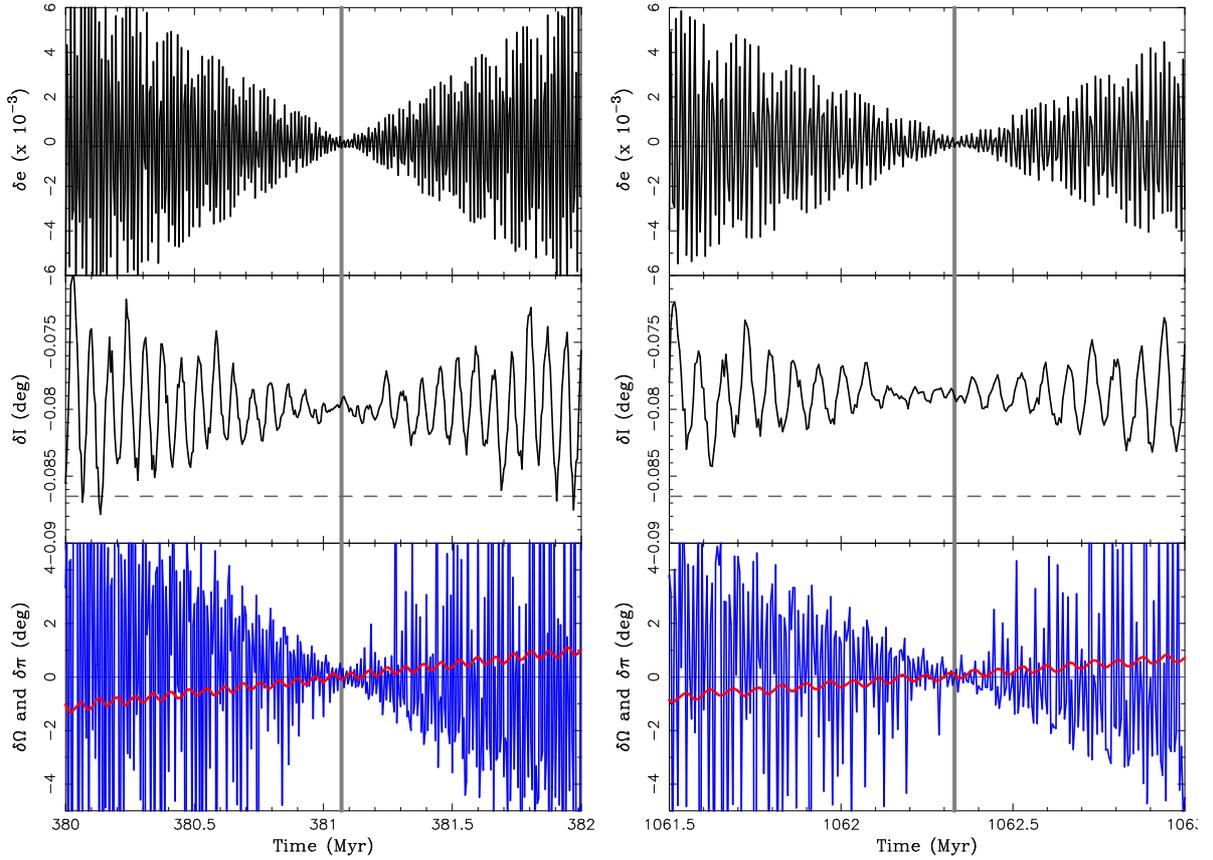}
\caption{Two examples of converging solutions between Yarkovsky
 clones of (258656) 2002~ES$_{76}$ and 2013~CC$_{41}$: left panels at $\simeq
 381.07$~Mya, right panels at $\simeq 1062.33$~Mya (gray vertical
 lines show the nominal convergence epochs). Each of the panels
 shows the differences between the osculating orbital elements of
 the clones: eccentricity (top), inclination (middle), and longitude
 of node (red) and perihelion (blue; bottom). The secular angles
 $\Omega$ and $\varpi$ converge to better than $0.004^\circ$,
 corresponding to a negligible value of the target function $\delta
 V\leq 0.04$ m~s$^{-1}$ (see Eq.~\ref{eq1}). Differences in $e$
 and $I$ are relatively larger, namely $\delta e \simeq 7.5\times
 10^{-5}$ and $\delta I\simeq 0.079^\circ$ (left), resp. $\delta
 I\simeq 0.078^\circ$ (right). The dashed horizontal lines show
 the differences  
 between the respective proper elements of (258656) 2002~ES$_{76}$ and
 2013~CC$_{41}$. Note the ordinate of the middle panel (inclination)
 which is offset from zero.}
\label{f10}
\end{figure*}

Figure~\ref{f9} shows the results from our Yarkovsky clone simulations. 
In contrast to the simulations where only the geometrical clones were 
used (Fig.~\ref{f8}), there are many more convergent solutions, 
starting from $360$~Mya. The reason is illustrated in 
the bottom
panel of Fig.~\ref{f9}, which shows the divergence of the osculating 
longitude of the ascending node between the nominal orbit (no Yarkovsky effect) 
and the clone with the maximum positive drift-rate $(da/dt)_{\rm max}$ 
of (258656) 2002~ES$_{76}$. Clones with smaller $da/dt$ values have nodal differences 
smaller than the signal seen in Fig.~\ref{f9}, proportionally to 
their $\cos\gamma$ value. 

The nodal differences between various clones are now much larger, reaching the maximum possible value of
$360^\circ$ after at $\simeq 1.1$~Gya. The nodal difference to the nominal orbit of 
the clone with the maximum negative drift-rate value is about the 
same but negative. This is because $\delta \Omega$ now propagates 
nearly quadratically in time as opposed to the quasi-linear trend
for the geometrical clones. Such a quadratic trend in node propagation 
is characteristic of 
Yarkovsky studies of asteroids \citep[e.g.,][]{vn08}. In that case, the phenomenon was easily 
associated with the principal dynamical perturbation produced by the 
Yarkovsky effect, namely the secular drift in semimajor axis. As a 
result, the semimajor axis dependence of the $s$ frequency produces, 
after a straightforward integration, a quadratic-in-time drift of 
the node. In our case of Jovian Trojans, the effects are slightly 
subtler. This is because, in spite of a permanent transverse perturbing 
acceleration in orbits of the clones, their semimajor axis does not 
show any constant drift in time due to the resonant locking inherent to their presence in the Trojan population. 
However, other elements --eccentricity and inclination-- do display
such a secular drift, as previously found in \citet{wh17} and \citet{hetal19}. As the $s$ frequency is also 
dependent on these values, it still displays 
a linear change as a function of 
time, explaining the quadratic effect in node seen in the Fig.~\ref{f9}.

Returning to the pattern in the distribution of converging solutions 
seen in Fig.~\ref{f9}, we note their clustering near epochs 
when $\delta\Omega$ of the (258656) 2002~ES$_{76}$ and 2013~CC$_{41}$ nominal 
orbits has been found to reach zero (the grey line in the top panels). This 
is to be expected, since 
the nodal difference exhibits the most 
stable evolution in time. Therefore, when nominal orbits of 
the two Trojans have large $\delta\Omega$ values, the clones 
will also follow the same pattern. This conclusion will, however, weaken 
further into the past because of the clone nodal divergence 
discussed above. As a result, beyond $\sim$ one Gyr into the past, the 
solution distribution spreads more in time. This is because 
specific clone combinations may now satisfy more easily 
our convergence conditions. Additionally, convergent solutions 
cluster in peaks separated by about $19$~Myr, rather than 
exhibiting a continuous distribution about the $\delta\Omega\simeq 
0$ nodal conditions. This is due to the $\delta\varpi\simeq 0$ 
perihelion condition also facilitating the convergence criteria 
we adopted. 

The middle panel in Fig.~\ref{f9} shows the 
statistical distribution of the number of converging solutions 
for a sub-sample of cases in which clones of (258656) 2002~ES$_{76}$ and 2013~CC$_{41}$ 
both have the same sign of the associated $da/dt$ drift. Translated 
using the diurnal Yarkovsky theory, this also implies that the two 
clones have the same sense of rotation: either both prograde, or both retrograde. 
The proposed formation mechanisms for this 
pair, namely a binary split or rotation fission, would both 
predict this property. There are obviously fewer
solutions found, but the general pattern of their distribution is 
about the same as in the general case when all clones are 
taken into account.

Figure~\ref{f10} shows the conditions at convergence for two pairs of
the Yarkovsky clones of (258656) 2002~ES$_{76}$ and 2013~CC$_{41}$: the left panels
at the most recent possible cluster of solutions in the past (namely at $\simeq 381.07$~Mya), whilst 
the right panel shows the cluster at an epoch which is
more distant in the past by two cycles of the differential motion
of their orbital nodes (namely at $\simeq 1062.33$~Mya). In
general, the quality of the convergence is similar, including those solutions beyond $1$~Gya. In both cases, the formal
convergence of the secular angles is better than $0.004^\circ$.

When inserted into Eq.~(\ref{eq1}), the equivalent velocity
difference is negligibly small $\delta V\leq 0.04$ m~s$^{-1}$.
At the convergence epoch, the osculating eccentricity values are
also satisfactorily close to each other, namely $\delta e\simeq 7.5\times 10^{-5}$. Using the Gauss equations \citep[e.g.,][]{nv06},
we 
estimate that this tiny eccentricity difference corresponds
to an orbital velocity change smaller than $1$ m~s$^{-1}$ in a
statistical sense. This change is actually smaller than the difference in
proper eccentricity values of (258656) 2002~ES$_{76}$ and 2013~CC$_{41}$. The
inclination convergence turns out to be the most troublesome
element of the simulation: the persisting differences of $\simeq
0.085^\circ$ statistically correspond to a velocity change of
$\simeq 25$ m~s$^{-1}$. Such a difference in the osculating values
of inclinations corresponds to the difference of their proper
values. In contrast, the acceptable true separation velocity of
the objects should be a fraction of the escape velocity from the
effective parent body. With its size of $\simeq 9$~km, the ideal
condition of the separation in this pair would require a velocity
difference of $\leq 4$ m~s$^{-1}$. The inclination difference at
converging solutions is therefore nearly an order of magnitude larger. 

One possibility to explain this mismatch may be related to our approximation 
of the Yarkovsky effect. By representing it using the transverse acceleration 
only, the inclination is not perturbed. In fact, a complete model of the thermal 
accelerations may admit an out-of-plane component, provided that the obliquities of the components of the pair are not extreme \citep[e.g.,][]{v98}. However, to fully use such a model, we would 
need to sample a multi-parametric space of possible spin orientations and physical 
parameters for Yarkovsky clones, an effort which is postponed to further studies. 

An alternative dynamical mechanism, that has not been included in our simulations, 
consists of perturbations
from the largest Trojans in the L$_4$ swarm. As an example, we consider the influence 624 Hektor, whose mass is estimated to be $\simeq 10^{17}$~kg 
\citep[e.g.,][]{c12}, about $10^{-4}$ of the mass of dwarf-planet 1~Ceres. 
\citet{netal02} found that, statistically, the mean perturbation of the orbital 
inclination produced by Ceres in the inner and middle parts of the Main
belt is $\simeq 1.5^\circ$ in $4$~Gyr. Assuming the effect scales with the 
square root of the perturber mass, we estimate that the  
approximate effect of Hektor on small L$_4$ Trojans would be 
$\simeq 0.015^\circ$ over $4$~Gyr, in a statistical 
sense. Therefore, at least a part of the inclination mismatch reported above 
could well be due to the ongoing scattering influence of the most massive Trojans.

\section{Formation of the Trojan pair} \label{disc}
We now briefly discuss possible formation processes for the (258656) 2002~ES$_{76}$--2013~CC$_{41}$ pair.
In principle, these mechanisms coincide with the suggestions outlined in Sec.~6 of \citet{vn08}. 
Building on that work, we will skip for now 
the possibility that these two Trojans are the two largest objects in a compact, collisionally-born family. Given their comparable size,
the collision required to form such a family must have been super-catastrophic, with many kilometer size fragments 
created and dominating the mass. Without information about them, it is hard to say 
anything more about the putative collision conditions, including the probability of such a collision actually having occurred.

\subsection{Collisional dissociation of a synchronous binary} \label{dissBinary}
The first possible origin for the (258656) 2002~ES$_{76}$--2013~CC$_{41}$ pair 
consists of a model, in which the two objects
were formerly components in a binary system which underwent some kind of
instability. We assume that the instability was not of a dynamical origin. Indeed,
even if formed by gravitational collapse, the initial angular momentum of
the binary would exceed that of a critically rotating single body of an equivalent
mass by a factor of $\simeq (3-10)$ \citep{Nesvorny2019TNBinaries}. This is not sufficient 
to drive tidal evolution, whilst conserving angular momentum, to the stability limit at about half of
the Hill sphere, even in 
the Trojan zone. The limiting configuration would
require angular momentum at least twice as large. Additionally, time constraints
may prevent evolution to such large separations within $\leq 4.5$~Gyr. Therefore,
the nature of the parent binary instability must be different. We assume instead that this instability was
triggered by a gentle-enough impact on one of the components.
We leave aside other possibilities, such as binary instability produced during a close three body encounter with a massive Trojan \citep{Agnor2006TritonCapture, nv19},for future investigations, once the mechanisms are better understood in the Jovian Trojan population.

\begin{figure}
 \includegraphics[width=\columnwidth]{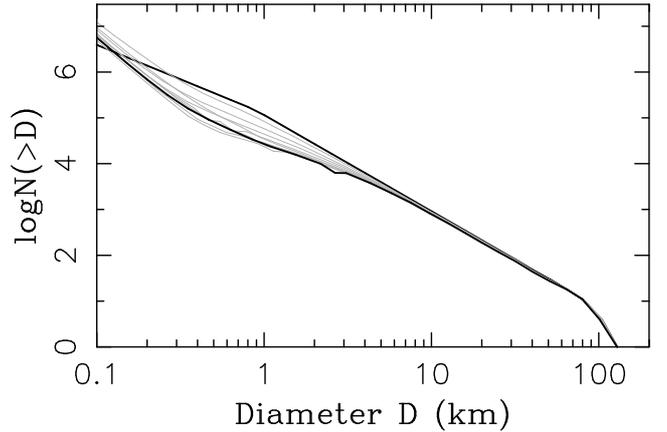}
\caption{The effects of collisional grinding on the cumulative size distribution
 of Jovian Trojans. The upper bold line is the initial distribution. The lower
 bold line is the size distribution at $T=4.5$~Gyr. The gray lines show the
 changing size distribution in $500$~Myr intervals. The dip in the final
 distribution near $D=0.5$~km is produced by the strength-to-gravity transition
 of the disruption law.}
 \label{f11}
\end{figure}

Let us start the likelihood analysis of the formation of the (258656) 2002~ES$_{76}$--2013~CC$_{41}$ 
pair via the sub-critical impact dissociation of a previously existing synchronous binary
with a very simple, order-of-magnitude estimate. Assume that the needed imparted 
velocity by the impact onto a $\simeq 7$~km size component in the binary is 
about $1$ m~s$^{-1}$. Then, using the simple formulation in \citet{netal11}, a 
projectile of $\simeq 0.53$~km size is required. The characteristic impact 
velocity assumed was $V_{\rm imp}\simeq 4.6$ km~s$^{-1}$ \citep{davaiii}. The 
Trojan population contains very approximately $N\simeq 400 000$ such objects 
\citep[e.g.,][and Fig.~\ref{f11}]{wb15,emeaiv}.

Using the mean impact probability $p_i\simeq 7\times 10^{-18}$ km$^{-2}$~yr$^{-1}$ 
\citep[e.g.,][]{davaiii}, we can therefore
estimate the order-of-magnitude likelihood
that such an event would occur within a timeframe of $T\simeq 4.5$~Gyr, namely $p_i R^2 N T
\simeq 0.15$ (here $R=3.5$~km is the radius of the target body). This suggests that every
such binary implanted to the Trojan population has a non-negligible (15\%) chance to be split via this process. 
Assuming that, initially, at least hundreds of binaries were captured intact to the Trojan population, 
a non-negligible number of Trojan pairs might have been created over the age of the Solar system. 
Obviously, in many cases, our ability to identify the pair produced in this manner is low, due to
unsuitable locations in the Trojan orbital phase space. Nonetheless, this result suggest that sufficiently many such pairs could be produced that future study might well reveal several more. 

We now substantiate this order-of-magnitude estimate using a more involved numerical 
simulation. As outlined above, the mutual orbit of a binary can be affected by 
small impacts on to its components. The binary may become unbound if the velocity 
change imparted by an impact exceeds binary's orbital speed $\sim 0.2-2$ m~s$^{-1}$ 
for bodies with $D\simeq 7$~km \citep{pm04}. 

We investigate this process with the previously developed collisional code
\citep[][]{metal09,netal11}. The code, known as {\tt Boulder}, employs a statistical 
method to track the collisional fragmentation of planetesimal populations. A full 
description of the {\tt Boulder} code, tests, and various applications can be found 
in \citet{metal09}, \citet{letal09} and \citet{betal10}. The binary module in {\tt 
Boulder} accounts for small, non-disruptive impacts on binary components, and 
computes the binary orbit change depending on the linear momentum of impactors 
\citep[see][]{netal11,nv19}.

We account for impacts over the life of the Solar system, $4.5$~Gyr. The captured population of Jovian Trojans 
is assumed to be similar to the present population, for objects with large diameters. 
There are $\simeq 25$ Trojans with $D>100$~km. The population is assumed to follow a power law profile 
below $100$~km, with a cumulative index equal to $-2.1$ (Fig.~\ref{f11}). The intrinsic impact 
probability and impact velocity is the same as used for the order-of-magnitude 
estimate above. We adopt a standard disruption law for solid ice from \citet{ba99}. 
Fragments are generated according to the method described in \citet{metal09}. These 
rules are implemented in the {\tt Boulder} code, which is then used to determine 
the collisional survival of Trojan binaries \citep[e.g.,][]{netal18}.

Figure~\ref{f11} shows the evolution of the size distribution for the Jovian Trojans. The size 
distribution for $D>10$~km remains unchanged over $4.5$~Gyr, but below $D\simeq 
5$~km the slope becomes shallower. This is consistent with Jovian Trojan 
observations that detect a shallower slope for $D\simeq 3$~km \citep[e.g.,][]{wb15}. 
If this interpretation is correct, the slope should become steeper below approximately 
$500$~m, for bodies that are too faint to be detected from the ground using the current generation of observatories. The dip in the 
size distribution is produced by the transition from strength-to-gravity dominated 
branches of the disruption law \citep[e.g.,][]{netal18}.

\begin{figure}
 \includegraphics[width=\columnwidth]{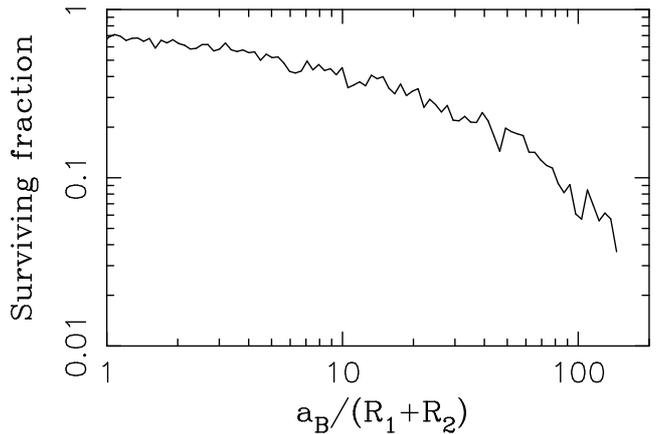}
\caption{The survival probability of binaries with (258656) 2002~ES$_{76}$--2013~CC$_{41}$
 components as a function of separation, here normalized to the sum of
 physical radii, $R_1+R_2$. The survival probability decreases with
 separation because wide binaries have smaller orbital speeds and are
 easier to dissolve by a small impact. For reference, the Hill radius
 $R_{\rm H}$ of a binary with (258656) 2002~ES$_{76}$--2013~CC$_{41}$ components,
 corresponding to mass $\sim 5\times 10^{17}$~g (for $1$ g~cm$^{-3}$
 density), is $R_{\rm H}\simeq 3,400$~km, or nearly $a_{\rm B}/(R_1+R_2)
 =500$.}
 \label{f12}
\end{figure}

We find that the survival chances of Trojan binaries are generally good, but drop significantly
when the binary separation approaches $0.5\,R_{\rm H}$ ($R_{\rm H}$ being the 
Hill sphere of gravitational influence, see Fig.~\ref{f12}). This is expected 
because binaries with semimajor axis $a_{\rm B}>0.5\,R_{\rm H}$ are dynamically
unstable \citep[e.g.,][]{pg12}. For a characteristic separation of $a_{\rm B}/ (R_1+R_2)\simeq 10-100$, 
where $a_{\rm B}$ is the binary semimajor axis and $R_1$ 
and $R_2$ are the binary component radii, consistent with the pair (258656) 2002~ES$_{76}$--2013~CC$_{41}$
($R_1+R_2=7.2$~km), which is quite common among equal-size binaries in the Edgeworth-Kuiper 
belt \citep[e.g.,][]{noll20}, the survival probability is $7-40$\%. There is plenty 
of room in this parameter space for Trojan pair formation by this mechanism. Assuming 
that the pair (258656) 2002~ES$_{76}$--2013~CC$_{41}$ is an impact-dissolved binary, we find that 
there should be $0.08-0.7$ surviving binaries for each pair such as 
(258656) 2002~ES$_{76}$--2013~CC$_{41}$. Given that the vast majority of Trojan pairs remain undetected 
(see the difficulties briefly outlined in the Appendix~\ref{addpairs}), the obvious 
implication is that there should also be several equal-size binaries among Jovian Trojans in this size range.

\subsection{Rotational fission of a parent object} \label{rotFission}
An alternative formation mechanism that could explain the observed properties of the
(258656) 2002~ES$_{76}$--2013~CC$_{41}$ pair is that they might be the result of the
rotational fission of their common parent object \citep[this is indeed the favorite mechanism for asteroid pair formation in the main belt; e.g.,][]{petal10}. The most
probable driving process for such a fission event
is the Yarkovsky-O'Keefe-Radzievski-Paddack (YORP)
effect, a radiative torque resulting from the combination of
reflected and thermally emitted radiation by the surface
\citep[being thus a complementary phenomenon to the Yarkovsky effect; e.g.,][]{betal06,vokaiv}.
The YORP effect is able to constantly accelerate an asteroid's rotation up to speeds that meet the requisite 
conditions to cause the object to fission
. The rotation frequency change ${\dot\omega}$ satisfies general scaling
properties, such that ${\dot\omega} \propto 1/[\rho\,(a D)^2]$, where $\rho$ is the bulk
density, $a$ the orbital semimajor axis and $D$ the size. However, the problematic part
of the YORP effect, unlike the Yarkovsky effect, is its large sensitivity to details of
the surface roughness. For that reason it is troublesome to determine the exact value of
the strength of the YORP effect for a given object,
and we must satisfy ourselves with an order-of-magnitude estimate in our case.

If we were to determine the doubling timescale  $\tau_{\rm YORP} =\omega/{\dot\omega}$
\citep[sometimes also the YORP cycle timescale; e.g.,][]{r00}, 
it would be reasonable to use the YORP detection
of the small near-Earth asteroid (101955) Bennu as a template, as we did above for the
Yarkovsky effect in section~\ref{yarko}. (101955) Bennu has $\tau_{\rm YORP}\simeq 1.5$~Myr
\citep[e.g.,][]{bennuyorp}. Adopting plainly the scaling $\tau_{\rm YORP} \propto \rho\,(a D)^2/P$
(with $P$ being the rotation period), we obtain $\tau_{\rm YORP}\simeq 11$~Gyr for a $D
\simeq 9$~km Trojan, the estimated size of a putative parent object of the (258656) 2002~ES$_{76}$--
2013~CC$_{41}$ pair. Note that $\tau_{\rm YORP}$ provides an estimate of a timescale for
doubling $\omega$, as an example changing rotation period from $5$~hr to $2.5$~hr, an
approximate fission limit for a large internal strength Trojan model. Another $\tau_{\rm YORP}/2
\simeq 5.5$~Gyr time would be needed if the initial rotation period of the parent object was
$10$~hr. This shorter timescale would also be an appropriate estimate to reach the fission
limit at a longer period of $\simeq 5$~hr when the internal strength and bulk densities are
low \citep[e.g.,][]{fetal15,setal17}. 

If, however, we were to consider the results from numerical
simulations of the YORP effect for a large statistical sample of Gaussian-sphere
shapes \citet{cv04}, which obtained $\tau_{\rm YORP}\simeq 15$~Myr for a typical main belt S-type asteroid
of a $2$~km size, we would have $\tau_{\rm YORP}\simeq 1.5$~Gyr for changing the parent object
period from $5$ to $2.5$~hr. Whilst these results are known to typically overestimate the
strength of the YORP effect by a factor of $3-5$, when compared to detections of the YORP
effect for small near-Earth asteroids, we nonetheless get a timescale shorter by a factor 2 to 3 than
for the Bennu case. The takeaway message is that the estimate of the YORP doubling
timescale prior the fission of the putative parent object of the (258656) 2002~ES$_{76}$ and 2013~CC$_{41}$
pair is very uncertain, with values ranging possibly from $2$~Gyr to some $12$~Gyr. 

Taken at a face value, the smaller values in this interval are plausible as an explanation for the origin of the 
pair when compared to the lifetime of the Solar system. It 
may not be surprising to find that some $D\simeq 9$~km Jupiter
Trojan objects undergo a rotational fission during their lifetime. However, a more detailed
inspection of the (258656) 2002~ES$_{76}$ and 2013~CC$_{41}$ parameters speaks against this possibility.
First, we note that the known rotation periods of Jovian Trojans rarely have values smaller
than $8-10$~hr \citep[e.g.,][]{fetal15,setal17,retal17}, which suggests in turn 
that more than one
$\tau_{\rm YORP}$ timescale would be needed to reach fission from a typical initial rotation
state (though, admittedly, these known data concern larger objects). More importantly, though,
we note that the absolute magnitude difference of (258656) 2002~ES$_{76}$ and 2013~CC$_{41}$ is $\simeq
(0.2-0.3)$, depending on the database used. This implies that the two objects are nearly of
the same size. \citet{petal10} argued that the typical conditions of fission mechanics
require at least $1$~magnitude difference between the two components in pair. This is
because some degree of size disparity is needed to make the two components separate
onto distinct heliocentric orbits. Whilst exceptions have been found to this guideline \citep[see e.g.][]{petal19},
the majority of the known asteroid pairs, more than $90$\%, satisfy this condition of having a large enough
magnitude disparity. The components in the (258656) 2002~ES$_{76}$ -- 2013~CC$_{41}$ pair violate this rule and
would require special conditions for their separation to feasibly be the result of rotational fission.

\section{Conclusions} \label{Concl}
In this work, we identified the first potential dynamical pair in the Jovian Trojan
population. In particular, we analysed the distribution of Trojans in 
their proper orbital element space. Using information about the local density of
objects, we also assessed the statistical significance of the proximity of potential couples.
This procedure lead us to select a pair of bodies, (258656) 2002~ES$_{76}$ and 2013~CC$_{41}$, in 
the L4 swarm as a potential candidate pair. Interestingly, this suggested pair is located very
close to the L4 Lagrange point, with low proper elements, semimajor axis ($da_{\rm P}$),
eccentricity $e_{\rm P}$ and sine of inclination ($\sin I_{\rm P}$) values. Finally,
as part of our effort, we developed an up-to-date, highly accurate set of proper
elements for the all Jovian Trojans, which we have made publicly available (Appendix~\ref{PropElements}). 

In order to further investigate the selected pair, we ran a series of $n$-body simulations,
which were used to look for past convergences in the osculating nodal ($\delta\varpi$) and perihelion
longitude ($\delta\Omega$) value for the two objects, whilst ensuring that, at the time of such convergences
the differences in the osculating
eccentricity and inclination were also sufficiently small. Our simulations included
both geometric clones, created from the uncertainties in the orbital elements of the bodies, and
Yarkovsky clones, based on the estimated thermal accelerations that the two objects could experience, for a variety 
of realistic rotation rates. As a result, we obtained a statistical set of convergences, finding a larger pool of 
possibilities once the Yarkovsky clones were included. Our results reveal that the pair is at least $\simeq 360$~Myr old, but are compatible with the age being significantly 
older, potentially in the Gyr time scale. By finding such possible convergences, we increase the confidence that the
(258656) 2002~ES$_{76}$--2013~CC$_{41}$ couple is a legitimate pair.

We then considered the mechanisms by which the
(258656) 2002~ES$_{76}$--2013~CC$_{41}$ pair could have formed \citep[compared with][]{vn08}.
The pair is not associated with any known collisional family, and as such we do not favour the
possibility of the pair having been formed as a result of a catastrophic impact on a putative parent body
. The pair might have been formed through the
rotational fission of their parent Trojan, since, for certain initial conditions, the timescale for such an object to be spun-up by the YORP effect to the point that it undergoes fission
could be somewhat shorter than the age of the Solar system. However, this pair consists of two nearly
equal-sized components, whilst the vast majority of observed pairs formed by rotational
fission have a size ratio of at least $1.5$ \citep[see][]{petal10,petal19}. For that 
reason, we consider
that the pair most likely formed as a result of
the dissociation of an equal-size binary. We can confirm that such a scenario is indeed feasible
using an estimation of the binary survival rate in the size range of the (258656) 2002~ES$_{76}$--2013~CC$_{41}$ pair, $D\simeq 7$~km, over
$4.5$~Gyr, after implantation to the Trojan population early in Solar system's history. Statistically, this indicates
that there should be many such pairs within the Trojan population in this $5-10$~km size range. As the Rubin Observatory's Legacy
Survey of Space and Time (LSST) comes online, it is expected to discover many Jovian Trojans
in this size range \citep[e.g.,][]{Schwamb2018LSST}. As new Trojans are discovered, our results suggest that further pairs should be revealed.

The (258656) 2002~ES$_{76}$--2013~CC$_{41}$ pair provides an interesting clue to the past history of the 
Jovian Trojans, and the Solar system as a whole. So far, we know little beyond their
dynamical properties and size estimations. In particular, lightcurve analysis could assist in constraining the formation mechanism, as this would provide an estimate of the rotational periods of the two objects. Due to their small size, and dark albedo,
the objects have relatively low apparent magnitudes, at best $\simeq 20.5$ magnitude in
visible band. In order to further characterize these objects, observations using
large Earth-based facilities, such as the SUBARU \citep{Kashikawa2002SubaruFOCAS} or
Keck \citep{Oke1995KeckLRIS} telescopes, will be required. These objects would also
benefit from future observations using the James Web \citep[JWST,][]{Rivkin2016JWSTAsteroids} and Nancy
Grace Roman Space Telescopes \citep[RST, formerly known as WFIRST,][]{Holler2018SolarSysWFIRST}.
Time on these telescopes is competitive, but we recommend proposals for observations of
(258656) 2002~ES$_{76}$ and 2013~CC$_{41}$ be selected to further extend our understanding of this 
interesting pair of Trojans.

\section*{Data Availability}
The database of Jovian Trojan proper elements is accessible at \url{https://sirrah.troja.mff.cuni.cz/~mira/mp/trojans\_hildas/}, and is available for community use. See Appendix \ref{PropElements} for details.

\section*{Acknowledgements}

T.R.H was supported by the Australian Government Research Training Program
Scholarship. The work of D.V. an M.B. was partially supported by the Czech
Science Foundation (grant 18-06083S). This research has made use of NASA
Astrophysics Data System Bibliographic Services. We thank Dr. Romina Di Sisto for their valuable review of this manuscript.

We dedicate this paper to the
memory of Andrea Milani and Paolo Farinella, who were the first to propose the
idea of a genetically connected pair of objects in the Jovian Trojan population
\citep{m93}.



\bibliographystyle{mnras}
\bibliography{pairs} 



\appendix

\section{Determination of the Jovian Trojan proper elements} \label{PropElements}
Here we briefly review our approach to compute synthetic proper elements for the 
currently known Jovian Trojan population. The method is based on \citet{m93}, see also
\citet{br11}, though we needed several modifications of the digital filters in
order to stabilize determination of the proper elements for Trojans having very
small libration amplitude. Our dynamical model included four giant planets, with barycentric corrections to compensate for the indirect perturbations
for terrestrial planets. This arrangement suitably speeds up computations when
dealing with the whole population of many thousands of Trojans. Nevertheless, we
also checked validity of our results using a dynamical model including also the
terrestrial planets in a full-fledged manner for a sub-sample of Trojans (notably
the low-$\delta V_{\rm P}$ that is of interest here). No significant differences
were observed. The initial planetary state vectors were taken from the JPL ephemerides
and those of the Trojans from the {\tt AstOrb} catalogue as of April~28, 2020, from
which their population was also identified.

We used well tested numerical package {\tt swift} \citep[e.g.,][]{ld94}, specifically
the MVS2 symplectic integrator \citep[e.g.,][]{lr01}, that we adapted for our
application in several ways. The most important was an implementation
of digital filters, helping us to eliminate short-period and forced terms from
osculating orbital elements, necessary for identification of the proper terms.
Due to the absence of the direct perturbations from the terrestrial planets, we
can allow a fixed integration timestep of $0.25$~yr. The input sampling into the
filtering routines was $1$~yr. We used a sequence of the convolution (Kaiser-window)
filters A A B \citep[e.g.,][]{qetal91} with decimation factors 10 10 3, which were
applied to the non-singular elements $z=k+\imath h=e\,\exp(\imath\varpi)$ and
$\zeta=q+\imath p= \sin I\,\exp(\imath\Omega)$. The intermediate time window for
this filtering procedure and output timestep was $300$~yr. At this stage,
the short-period terms with periods comparable to planetary orbital periods or
the libration period were efficiently suppressed from the resulting mean values
${\bar z}$ and ${\bar \zeta}$ of eccentricity and inclination variables. We then
accumulated batches of 2048 values of ${\bar z}$ and ${\bar \zeta}$, and applied
Fourier transformation \citep[in particular the FMFT method from][]{sn96}, on the output. After
rejecting signal associated with forced planetary frequencies (such as $g_5$,
$g_6$ or $s_6$ to recall the principal ones), we were left with the proper values
$e_{\rm P}$ for the eccentricity and $I_{\rm P}$ for the inclination as the amplitude
of the remaining dominant terms. Our simulation spanned the total of $30$~Myr, and
we computed proper elements in the $\simeq 600$~kyr window described above many
times over intervals with $100$~kyr shift in their origin. This way we had a series
of many tens of proper element realizations, allowing to access their stability
and compute their mean and variance. We also observed that the series of individual
$e_{\rm P}$ and $I_{\rm P}$ still contained long-period signal (periods $> 1$~Myr),
which in future studies may call for extension of integration windows. At this moment,
we however, satisfied ourselves with our set-up. We also used the above outlined
procedure for the semimajor axis $a$, but instead of applying FMFT on its mean values
we simply computed its mean value ${\bar a}$ over a $1$~Myr interval. This helps us
to determine semimajor axis value of the libration center for a given Trojan orbit.

In order to obtain a reliable information about a stable libration amplitude we need
to apply a different method that has been implemented in our code in parallel to
computation of $e_{\rm P}$ and $I_{\rm P}$. This is because the corresponding libration
frequency is fast, $f \simeq 2.434$ deg~yr$^{-1}$ and $360^\circ/f \simeq 148$~yr, and
must not be under-sampled. A delicate issue consists of the fact that, at the same
time, one has to suppress terms with period even shorter than the libration period,
namely those which are related to orbital periods of giant planets (principally
Jupiter, $\simeq 11.86$~yr). We thus applied convolution filters B B, with decimation
factors 3 3, to the osculating values of the semimajor axis $a$ and the longitude
difference $\lambda-\lambda'$ (the orbital elements labeled with prime correspond
to Jupiter), a resonant argument of the Trojan tadpole motion. These intermediate
(mean) values of $a$ and $\lambda-\lambda'$ are computed with a $9$~yr cadence. In
the next step, the intermediate $a-a'$ were fitted by a straight line and the
constant term $a_0$ was subtracted. In the same way, the intermediate angle $\phi
= \lambda-\lambda'-\chi$, where $\chi = \pm 60^\circ$ depending on the L4 and L5
libration points, was fitted by a straight line and the constant term $\phi_0$
was subtracted. Effectively, after subtractions of the mean values was done, the
tadpole motion around the Lagrange point centers in these rescaled, zero-averages
$a-a'$ vs $\phi$ coordinates is centered at the origin. Consequently, the polar
angle $\psi$ defined as \citep[see, e.g.,][$a$ and $a'$ in au]{m93}
\begin{equation}
 \psi = \arctan\left(\frac{a-a'}{0.2783\,\phi}\right)
\end{equation}
can be unfolded by $360^\circ$, fitted by a straight line, with the slope defining the
libration frequency $f$. The libration amplitudes $da_{\rm P}$ (in au) and $D$ (in deg)
are computed by the Fourier transform as amplitudes of spectral terms with frequency
$f$. This second step uses a $1$~kyr cadence. Finally, we apply another averaging of
$da_{\rm P}$ and $D$ values, defined on a simple running window with the output time
step of $1$~Myr. Both $da_{\rm P}$ and $D$ may be considered as the third proper orbital
element alongside of $e_{\rm P}$ and $I_{\rm P}$.
\begin{figure}
 \includegraphics[width=\columnwidth]{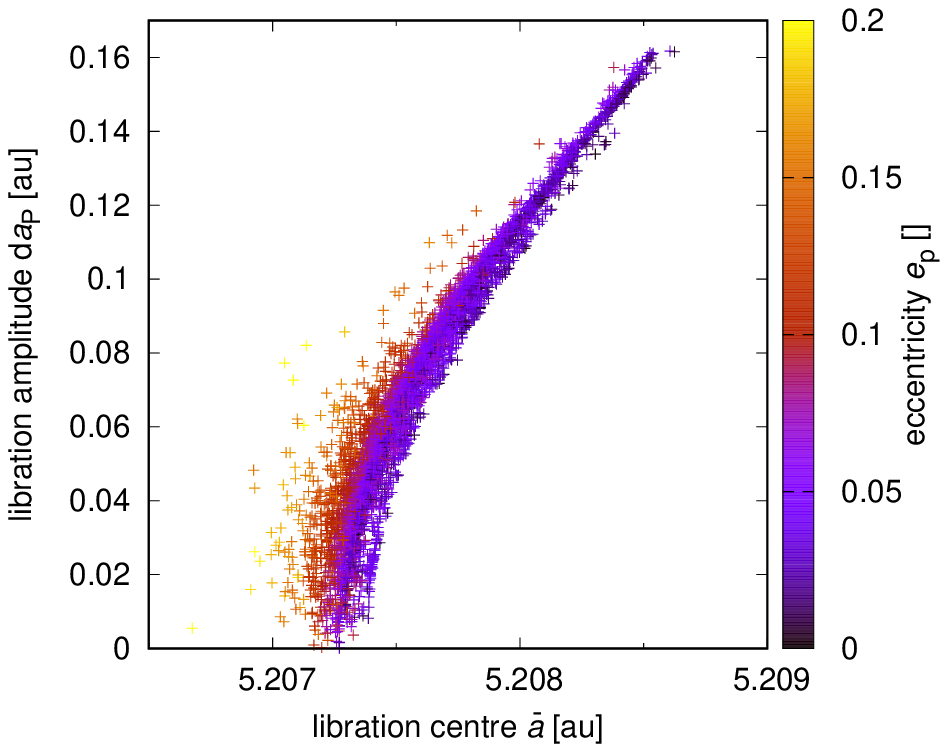}
 \caption{Libration amplitude $da_{\rm P}$ vs libration center $\bar a$ for Trojans
  in the L4 region. Colour corresponds to the proper eccentricity $e_{\rm P}$. The
  dependence of $\bar a(da_{\rm P},e_{\rm P})$ is systematic, indicating a functional
  dependence.}
 \label{f13}
\end{figure}

We note that the value of libration center ${\bar a}$ is not universal for all Trojans.
Instead, its value functionally depends on the proper elements $(da_{\rm P},e_{\rm P},I_{\rm P})$
or $(D,e_{\rm P},I_{\rm P})$, see Fig.~\ref{f13}. Some authors \citep[e.g.,][]{br11,retal16}
thus define an alternative set of proper elements $(a_{\rm P}={\bar a}+da_{\rm P},e_{\rm P},
I_{\rm P})$.

We determined the above-introduced parameters, including different variants of orbital proper
values and their uncertainty, for $7328$ Jovian Trojans, population as of April~2020. These
data can be found on \url{https://sirrah.troja.mff.cuni.cz/~mira/mp/trojans\_hildas/}.

\section{Are there more low-$\delta V_{\rm P}$ couples?} \label{addpairs}
As also suggested by data in Fig.~\ref{f5}, the brief answer to the topic of
this Appendix is probably positive, but a full analysis if this issue is left to
the future work. Here we only restrict ourselves to illustrate difficulties one
would quickly face in attempting to prove the past orbital convergence on a Gyr
timescales for most of the candidates.

Let us consider another low-$\delta V_{\rm P}$ candidate couple characterized by small
values of proper orbital elements $(da_{\rm P},e_{\rm P},\sin I_{\rm P})$, which helps
to minimize the unrelated background Trojan population (section~\ref{sel}). Staying
near the L4 libration point, we find 219902 (2002~EG$_{134}$) and 432271 (2009~SH$_{76}$)
at $\delta V_{\rm P}\simeq 4.9$ m~s$^{-1}$ distance. This couple has also appreciably small
probability $p\simeq 1.5\times 10^{-6}$ to be a random fluke and it has been highlighted
by a green circle in Fig.~\ref{f5}. The proper elements read $da_{\rm P}
\simeq (7.0372\pm 0.0004)\times 10^{-3}$~au, $e_{\rm P}\simeq (3.87534\pm 0.00004)\times
10^{-2}$ and $\sin I_{\rm P}\simeq (9.496\pm 0.003)\times 10^{-2}$ for (219902) 2002~EG$_{134}$, and
$da_{\rm P}\simeq (7.2950\pm 0.0006) \times 10^{-3}$~au, $e_{\rm P}\simeq (3.87652\pm
0.00004)\times 10^{-2}$ and $\sin I_{\rm P}\simeq (9.469\pm 0.002)\times 10^{-2}$ for
(432271) 2009~SH$_{76}$ (for reference, we again mention their quite small libration amplitudes
$1.44^\circ$, resp. $1.48^\circ$). This is a configuration reminiscent of the
(258656) 2002~ES$_{76}$-2013~CC$_{41}$ case, though each of the three proper elements is slightly larger
now. The relative velocity $\delta V_{\rm P}$ is again entirely dominated by the proper
inclination difference, this time somewhat smaller than in the (258656) 2002~ES$_{76}$-2013~CC$_{41}$
case (only $\simeq 0.015^\circ$). Assuming geometric albedo value $0.075$, we obtain
sizes of $\simeq 12.8$~km and $\simeq (7.3-8.1)$~km for (219902) 2002~EG$_{134}$ and (432271) 2009~SH$_{76}$, considering
absolute magnitude values from the major three small-body ephemerides sites as above.
While little larger, it still places this couple into the same category of very small
Trojans as (258656) 2002~ES$_{76}$-2013~CC$_{41}$.
\begin{figure}
 \includegraphics[width=\columnwidth]{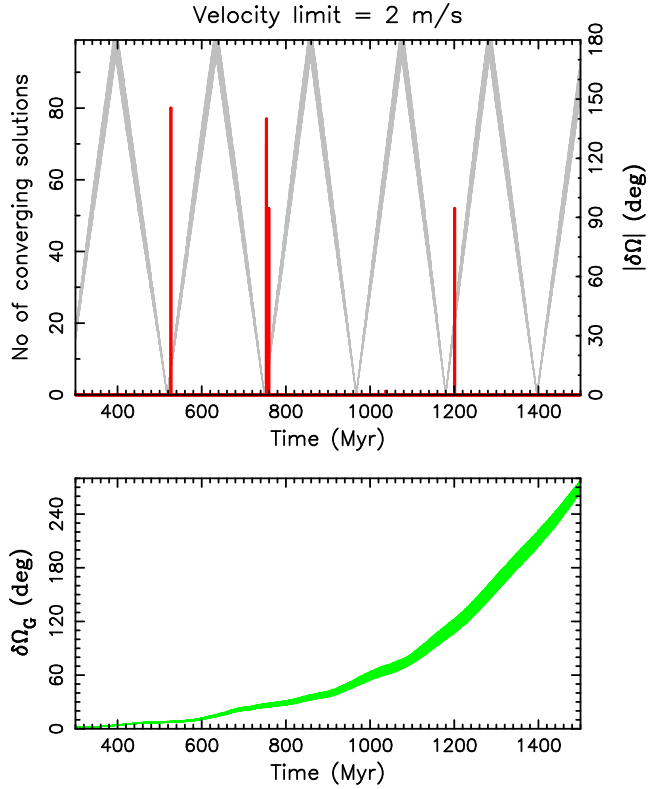}
 \caption{The same as Fig.~\ref{f8} but for the (219902) 2002~EG$_{134}$--(432271) 2009~SH$_{76}$ couple of Trojans: past
  orbital histories of nominal orbits and 20 geometrical clones each compared every
  $500$~yr and convergent solutions within $\delta V\leq 2$ m~s$^{-1}$ limit combined in
  $50$~kyr bins. Top panel gives the number of solutions for all possible combinations
  of clones (red histogram). The gray line gives $|\delta \Omega|$ of the nominal orbit of (219902) 2002~EG$_{134}$ and
  (432271) 2009~SH$_{76}$ (see also the right ordinate). The green line at the bottom
  panel shows the maximum difference in longitude of ascending node between the clones of (219902) 2002~EG$_{134}$
  and the longitude of ascending node of its nominal orbit, compared with the same information for
  (258656) 2002~ES$_{76}$ given in Fig.~\ref{f8}).}
 \label{f14}
\end{figure}

We repeated the convergence experiment using geometrical clones from section~\ref{geom}.
In particular we considered nominal (best-fit) orbits of (219902) 2002~EG$_{134}$ and (432271) 2009~SH$_{76}$, and for each
of them we constructed 20 geometrical clone variants of the initial data at MJD58800 epoch.
We again used information from the {\tt AstDyS} website and noted that both initial orbits
of components in this possible couple have smaller uncertainties in all orbital elements
than the orbits of (258656) 2002~ES$_{76}$ and 2013~CC$_{41}$. This is because their longer observation arcs
and more data available for the orbit determination. We propagated these 42 (21+21) test
bodies backward in time to $1.5$~Gyr before present. Perturbations from all planets were
included and every $500$~yr configuration of the nominal orbits and accompanied clones for
the two bodies compared. A criterion for convergence included $\delta V\leq 2$ m~s$^{-1}$
from Eq.~(\ref{eq1}), and small eccentricity and inclination differences. In particular,
we required $\delta e\leq 10^{-4}$ and $\delta I\leq 0.029^\circ$. These values are only
slightly larger than the difference in the corresponding proper values and and each represent
a few meters per second contribution in (\ref{eq-d}).

Results are shown in Fig.~\ref{f14} which has the same structure as the Fig.~\ref{f8},
previously given for the (258656) 2002~ES$_{76}$ and 2013~CC$_{41}$ couple. The main take-away message is
in the bottom panel, which shows maximum nodal difference between clones of (219902) 2002~EG$_{134}$ and
its nominal orbit as a function of time to the past. The slope of the initially linear
trend (lasting approximately $50$~Myr) is simply given by maximum $\delta s$ proper frequency
among clones from the initial data difference. The non-linearity, which develops at later
epochs, is due to orbital long-term chaoticity. While for the (258656) 2002~ES$_{76}$ and 2013~CC$_{41}$
couple the chaotic effects were very minimum, the nodal difference between (258656) 2002~ES$_{76}$ clones
and the nominal orbit increased to only $\simeq 4^\circ$ in $1.5$~Gyr. At the end of our run the nodal difference expanded to $\simeq 260^\circ$. Given the very
limited number of clones we had, this works again identification of convergent solutions.
Note that beyond $\simeq 970$~Myr, where we would expect more convergent cases, we could
satisfy the convergence criteria of only few meters per second described above only
rarely. CPU-demanding effort with many more clones would be needed to achieve the desired
convergence limits.

We repeated the same experiment for several other candidate couples from the small-$\delta
V_{\rm P}$ sample, including the case of (215110) 1997~NO$_5$--2011~PU$_{15}$ (see Fig.~\ref{f2}), but
observed even faster onset of the clone diffusion in the Trojan orbital phase space. This
was due to their large $e_{\rm P}$ and/or $\sin I_{\rm P}$ values, as well as larger
libration amplitudes. Their systematic analysis is beyond the scope of this paper. 


\bsp	
\label{lastpage}
\end{document}